\documentclass{mn2e}
\usepackage{amsfonts}
\usepackage{amsmath}
\usepackage{amssymb}
\usepackage{graphicx}

\newcommand{\p}{\ensuremath{\partial}}
\newcommand{\del}{\ensuremath{\delta}}
\newcommand{\Del}{\ensuremath{\Delta}}

\newcommand{\sig}{\ensuremath{\sigma}}

\newcommand{\epc}{\ensuremath{\epsilon_{\times}}}
\newcommand{\Sc}{\ensuremath{S_{\times}}}
\newcommand{\So}{\ensuremath{S_{0}}}
\newcommand{\delc}{\ensuremath{\delta_{\rm c}}}
\newcommand{\delo}{\ensuremath{\delta_{0}}}
\newcommand{\delh}{\ensuremath{\delta_{\rm h}}}
\newcommand{\rhoh}{\ensuremath{\rho_{\rm h}}}
\newcommand{\dprbar}{\ensuremath{{\bar\delta^\prime}}}
\newcommand{\sigdpr}{\ensuremath{\bar\sigma}}
\newcommand{\dprbarsq}{\ensuremath{{\bar\delta^{\prime2}}}}
\newcommand{\sigdprsq}{\ensuremath{\bar\sigma^2}}
\newcommand{\dcr}{\ensuremath{\delta_{{\rm c}\times}}}

\newcommand{\Msun}{\ensuremath{M_{\odot}}}

\newcommand{\avg}[1]{\ensuremath{\left\langle \,#1\, \right\rangle}}
\newcommand{\etal}{et al.}

\newcommand{\dir}{\ensuremath{\delta_{\rm D}}}

\newcommand{\der}{\ensuremath{{\rm d}}}

\newcommand{\xb}{\ensuremath{\mathbf{x}}}
\newcommand{\yb}{\mathbf{y}}
\newcommand{\zb}{\mathbf{z}}

\newcommand{\cb}{\ensuremath{\mathbf{c}}}
\newcommand{\C}{\ensuremath{\mathbf{C}}}

\newcommand{\kb}{\ensuremath{\mathbf{k}}}

\newcommand{\erfc}[1]{\ensuremath{{\rm erfc}\left(#1\right)}}
\newcommand{\erf}[1]{\ensuremath{{\rm erf}\left(#1\right)}}

\newcommand{\eqn}[1]{equation~\eqref{#1}}
\newcommand{\eqns}[1]{equations~\eqref{#1}}
\newcommand{\fig}[1]{Figure~\ref{#1}}
\newcommand{\figs}[1]{Figures~\ref{#1}}
\newcommand{\ph}[1]{\phantom{#1}}

\newcommand{\be}{\begin{equation}}
\newcommand{\ee}{\end{equation}}
\newcommand{\Cal}[1]{\ensuremath{\mathcal{#1}}}

\title[Scale dependent bias]
      {Scale dependent halo bias in the excursion set approach} 

\author[M. Musso \etal]
{Marcello Musso$^{1}$\thanks{E-mail: marcello.musso@uclouvain.be}, Aseem
  Paranjape$^{2}$\thanks{E-mail: aparanja@ictp.it} \& Ravi K. Sheth$^{2,3}$\\  
 $^1$ CP3-IMP3, Universit\'e Catholique de Louvain, 2 Chemin du Cyclotron, 
  1348 Louvain-la-Neuve, Belgium\\
 $^2$ The Abdus Salam International Center for Theoretical Physics,
  Strada Costiera, 11, Trieste 34151, Italy\\
 $^3$ Center for Particle Cosmology, University of Pennsylvania, 
      209 S. 33rd St., Philadelphia, PA 19104, USA}
\begin{document}
\pagerange{\pageref{firstpage}--\pageref{lastpage}}

\maketitle 

\label{firstpage}

\begin{abstract}
\noindent 
If one accounts for correlations between scales, then nonlocal, $k$-dependent halo bias is part and parcel of the excursion set approach, and hence of halo model predictions for galaxy bias. We present an analysis that distinguishes between a number of different effects, each one of which contributes to scale-dependent bias in real space.  We show how to isolate these effects and remove the scale dependence, order by order, by cross-correlating the halo field with suitably transformed versions of the mass field.  These transformations may be thought of as simple one-point, two-scale measurements that allow one to estimate quantities which are usually constrained using $n$-point statistics.  As part of our analysis, we present a simple analytic approximation for the first crossing distribution of walks with correlated steps which are constrained to pass through a specified point, and demonstrate its accuracy.  Although we concentrate on nonlinear, nonlocal bias with respect to a Gaussian random field, we show how to generalize our analysis to more general fields.  

\end{abstract}


\section{Introduction}
\label{intro}

Galaxy clustering depends on galaxy type (Zehavi \etal\ 2011 and 
references therein).  Therefore, not all galaxies 
are fair tracers of the dark matter distribution.  Precise constraints 
on cosmological models require a good understanding of this galaxy 
bias (Sefusatti \etal\ 2006; More \etal\ 2012). In the simplest models, 
galaxies are linearly biased tracers (Kaiser 1984), but, even at the 
linear level, this bias may depend on physical scale or wavenumber $k$ 
(e.g. Desjacques et al. 2010; Matsubara 2011).  This scale-dependence, 
which is clearly detected in simulations of hierarchical clustering 
models (Sheth \& Tormen 1999; Smith et al. 2007; Manera et al. 2010), 
contains important information about the statistics of the initial 
fluctuation field, and the nature of gravity (Parfrey, Hui \& Sheth 2011; 
Lam \& Li 2012).

The most common galaxy bias model -- the local bias model -- assumes 
that the galaxy overdensity field $\delta_h(x)$ is a local, possibly 
nonlinear, monotonic, deterministic transformation of the dark matter 
field $\delta(x)$ at the same position (Fry \& Gazta\~naga 1993; 
Manera \& Gazta\~naga 2012; Pollack, Smith \& Porciani 2012; 
Chan \& Scoccimarro 2012).
Even in this case, there are a number of ways in which scale 
dependence can arise, even for the simplest case of Gaussian 
initial conditions and standard gravity.  
Since the measured bias will generally be a combination of all these 
effects, we present some ideas on how to disentagle them from one 
another.  

In general, of course, $\delta_h$ might depend on the value of 
$\delta$ at different locations, on its derivatives (Desjacques et al. 2010; 
Musso \& Sheth 2012), on other higher order statistics of the field 
(e.g. Sheth, Mo \& Tormen 2001; Sheth, Chan \& Scoccimarro 2012) 
at the same or at different positions, etc.; the dependence might 
even not be deterministic (e.g., Sheth \& Lemson 1999; Dekel \& Lahav 1999).
Our final goal is to present methods which are able to pinpoint this
relation even when the bias is nonlinear, nonlocal and stochastic.  

We study insights which arise from the simplest treatment of halo bias:  
that based on the excursion set approach (Press \& Schechter 1974).  
This approach maps the problem of counting the number of collapsed halos 
to that of the first crossing of a suitable threshold (the `barrier') 
by random walks in density generated by smoothing the initial matter 
density field using a sequence of filters of decreasing scales 
(Bond et al. 1991).  
In addition to depending on the `barrier' shape, the first crossing 
distribution also depends on how far from the `origin' the walks happen 
to be for the largest smoothing scale $S_0$.

Walks that do not start from the origin have modified first crossing 
distributions (Lacey \& Cole 1993). This introduces a dependence of 
the abundance of halos $1+\delta_h$ on the initial matter density 
field $\delta$ smoothed on the much larger scale $S_0$, and hence leads 
to a prediction for halo bias (Mo \& White 1996).  

The excursion set approach greatly simplifies when the smoothing filter 
is sharp in Fourier space, because in this case the steps in each walk 
are uncorrelated with each other.  Since most analyses to date have 
relied on this choice, we use it to illustrate many of our key points.  
E.g., if the bias is deterministic and nonlinear in real-space, it will 
be stochastic in $k$-space.  And, estimates of cross-correlations 
between the halo and mass fields depend on the assumed form of the 
probability distribution function of the mass: one must be careful to 
use the appropriate probability density function (pdf).  
One of the key insights of this paper is to 
show that suitably defined real-space cross-correlation measurements 
allow one to extract the different bias coefficients, order by order.  

Recently, however, there has been renewed interest in studying the 
effects of smoothing with more realistic filters such as the TopHat in 
real space or the Gaussian. The problem is complicated in this case by 
the presence of nontrivial correlations between the steps of the random 
walks (Peacock \& Heavens 1990; Bond \etal\ 1991), and a number of different approximations for the effect on the 
first crossing distribution have been introduced (Maggiore \& Riotto 2010; 
Paranjape, Lam \& Sheth 2012).  We show that the most accurate of these, 
due to Musso \& Sheth (2012), can be extended to provide a very 
accurate model for walks which do not start from the origin.  

We then show that correlations between steps generically introduce two additional sources of scale-dependent bias into the predictions.  One is relatively benign, and simply arises from the fact that the excursion set prediction is for a real-space quantity, but the halo bias in $N$-body simulations is typically measured in \emph{Fourier} space, through ratios of power spectra.  That this matters reiterates a point first made by Paranjape \& Sheth (2012), but it is easily accounted for by using a more appropriate normalization of the bias coefficients.  The second is more pernicious and is a genuinely new source of $k$-dependent bias (a point made in Musso \& Sheth 2012, but not studied further).  
Although this complicates discussion of scale-dependent bias, our method of measuring suitably defined real-space cross-correlations between the halo and mass fields can be used to extract the $k$-dependence of halo bias order by order. 

This paper is organised as follows.  Section~\ref{uncorr} briefly summarizes known excursion set results for uncorrelated steps, defines the halo bias factors as a ratio of real space measurements, derives their large scale limiting values, uses these to motivate a real-space cross-correlation measurement at finite scale which returns these limiting values, and quantifies the importance of computing averages over the correct ensemble.  

Section~\ref{corrsteps} extends these results to the case of correlated steps.  We first derive the conditional distribution $f(s|\delo,\So)$ that a walk crosses the barrier for the first time at scale $s$ having taken up the value $\delta_0$ at scale $S_0$, and demonstrate its accuracy by comparing with the results of a Monte Carlo treatment of the problem.  We then turn to the problem of halo bias, and highlight some important differences from the uncorrelated case:  the question of the correct pdf is shown to be much less important, whereas the scale dependence of bias becomes more dramatic.  
We discuss some of the implications of our analysis and conclude in section~\ref{conclusions}.
Appendix~\ref{app-details} collects proofs of some results quoted in the text, while Appendix~\ref{app-matsubara} connects the bias coefficients defined using cross-correlation measurements to other definitions in the literature.

Throughout we will present results for a constant barrier of height \delc. Moving barriers pose no conceptual difficulty for the first crossing distributions we are interested in. Also, while our analytical results are generally valid for any smoothing filter and power spectrum, for ease of implementation, the explicit comparisons with numerical solutions will use the Gaussian filter and a power law power spectrum. Again, we do not expect our final conclusions to depend on this choice.

\section{The excursion set approach:  Uncorrelated steps}
\label{uncorr}

The excursion set ansatz relates the number of halos in a mass range $(m,m+\der m)$ to the fraction $f(s)$ of walks that first cross the barrier in the scale range $(s,s+\der s)$ through the relation
\be
 \frac{m}{\bar\rho}\frac{{\rm d}n(m)}{{\rm d}m} {\rm d}m = f(s)\,{\rm d}s, 
 \label{excsetansatz}
\ee
where $s=s(m) \equiv \avg{\del^2(m)}$ is the variance of the matter density field smoothed on a Lagrangian length scale corresponding to mass $m$ and linearly extrapolated to present day, and $\bar\rho$ is the background density.  

In this approach, the influence of the underlying dark matter field 
on the abundance of halos of mass $m$ (i.e.~the bias) can be 
estimated from the fraction $f(s|\delo,\So)$ of walks that first 
cross the barrier at $s$ starting from some prescribed height 
$\delo$ on some prescribed scale $\So$, rather than from the origin 
(Mo \& White 1996; Sheth \& Tormen 1999). 
The mean number overdensity of halos can be defined as
\be
 \avg{1+\delh|\delo,\So} \equiv \frac{f(s|\delo,\So)}{f(s)} \, ,
 \label{delhgivendel0}
\ee
which is explicitly a prediction in real space, and valid on 
scale $S_0$ in the Lagrangian initial conditions.  

Typically, the bias is characterised by expanding the above 
expression in powers of \delo. 
The coefficients of this expansion will in general depend on \So\
(besides obviously depending on $s$).
Moreover, the evaluation of $f(s)$ and $f(s|\delo,\So)$ (and 
therefore of the bias coefficients) is rather different depending 
on whether or not the steps in the walk are correlated.  
In what follows, we elucidate the issue of the scale dependence
of the bias coefficients in the simpler case of walks with 
uncorrelated steps. We also argue that the same coefficients can 
be obtained as the mean value of the product of 
$\avg{1+\delh|\delo,\So}$ and polynomials in \delo, weighted by 
the probability distribution of \delo. 
This alternative definition as an expectation value will be more
suitable to be extended to the case of correlated steps (section~\ref{corrsteps}), and to
make contact with the definition of bias in generic models 
other than the excursion set approach (Appendix \ref{app-matsubara}).

\subsection{Large scale Lagrangian bias factors}
The conditional first crossing distribution of a constant barrier \delc\ for walks with uncorrelated steps is (Bond et al. 1991; Lacey \& Cole 1993)
\be
 f_{\rm u}(s|\delo,\So) = \frac1{\sqrt{2\pi}} \frac{\delc-\delo}{(s-\So)^{3/2}}
                         {\rm e}^{-(\delc-\delo)^2/2(s-\So)}\,,
\label{fc-LC}
\ee
(the subscript in $f_{\rm u}$ standing for ``uncorrelated''), where 
$\delc > \delo$ and $s > \So$.  The corresponding unconditional 
distribution is
 $sf_{\rm u}(s)=(2\pi)^{-1/2}\nu{\rm e}^{-\nu^2/2}$,
where $\nu^2\equiv\delc^2/s$. In this case, setting $\So=0$ and 
expanding around $\delo=0$ leads to (Mo \& White 1996; Mo, Jing \& White 1997)
\be
 \frac{f_{\rm u}(s|\delo,\So=0)}{f_{\rm u}(s)} 
  = 1+\sum_{n=1}^{\infty} \frac{\delo^n}{n!}b^{\rm u}_n(\nu)\,,
 \label{bias-trad-shk}
\ee
with the bias coefficients given by
\be
 \delc^n b^{\rm u}_n = \nu^{n-1}H_{n+1}(\nu)\,,
 \label{bias-shk}
\ee
where $H_m(x)={\rm e}^{x^2/2}(-d/dx)^m {\rm e}^{-x^2/2}$ are the ``probabilist's'' Hermite polynomials. For example, $n=1$ returns the familiar expression for the linear halo bias $b^{\rm u}_1 = (\nu^2-1)/\delc$.  Note that these coefficients are pure numbers, independent of wavenumber $k$, and (by definition) of \So.  It is these scale-independent numbers which are most often used to derive cosmological constraints.  (Of course, for non-negligible \So, the Taylor series expansion of \eqn{delhgivendel0} will yield bias coefficients that depend on \So, but this dependence is almost never calculated or used.)  Since the $\So\to 0$ limit of \eqn{fc-LC} corresponds to setting $\delc\to\delc - \delo$ in the unconditional crossing distribution, these $b_n$ are simply related to the $n$th derivative of $sf_{\rm u}(s)$ with respect to \delc.  This makes it easy to see why the Hermite polynomials feature so prominently in much of what follows.

\subsection{A weighted-average definition of bias}
\label{hermitebias}
If we ignore the fact that the conditional distribution in equation~(\ref{fc-LC}) should really have $\delo<\delc$, then it is easy to check that
\be
f_{\rm u}(s) = \int_{-\infty}^{\infty}\der\delo\,
              f_{\rm u}(s|\delo,\So)\, p_{\rm G}(\delo;\So).
\label{fc-shk-incorrect}
\ee
where $p_{\rm G}(\delo;\So)$ is a Gaussian distribution with zero mean and
variance \So. Although this result is formally correct, we argue in the next subsection that the appropriate distribution over which to average should not be a Gaussian (nor even a Gaussian chopped at $\delo>\delc$).  
But if we continue to ignore this detail, then we find 
\be
\int_{-\infty}^\infty\der\delo\,p_{\rm G}(\delo;\So) \frac{f_{\rm u}(s|\delo,\So)}{f_{\rm u}(s)}\delo\, = b^{\rm u}_1\,\So
\notag
\ee
(Desjacques et al. 2010), and more generally, the orthogonality of the Hermite polynomials implies (Appendix~\ref{app-shk}) that 
\be
\frac{\delc^n}{\So^{n/2}} \avg{\frac{f_{\rm u}(s|\delo,\So)}{f_{\rm u}(s)}\,H_n\left(\frac{\delo}{\sqrt{\So}}\right)} = \delc^n b^{\rm u}_n .
\label{bn-crosscorr-shk}
\ee
This exact result is remarkable because the left hand side involves
quantities for an arbitrary \So, whereas the right hand side, which is
independent of \So, is simply the $\So\to 0$ limit of the appropriate
bias coefficient.  This is not at all obvious if one had viewed the
local bias expansion as a formal Taylor series:  one would naively
have thought that, at the very least, the cross correlation
$\avg{(1+\delh)\delo}$ should involve the bias coefficients of all
(odd) orders (for a further discussion, see Frusciante \& Sheth 2012).

Strictly speaking, this is only a mathematical curiousity, since the conditional distribution $f_{\rm u}(s|\delo,\So)$ is formally zero for $\delo>\delc$, but the identity above holds only when (incorrectly) averaging the expression in \eqref{fc-LC} over the full (Gaussian) distribution of \delo.  However, if we forget for the moment about how the bias factors in equation~(\ref{bias-shk}) were determined, then the analysis above shows that the $\So\to 0$ limit of the bias coefficients can be recovered by cross-correlating the halo overdensity field with a suitably transformed version of the mass field (the transformation uses Hermite polynomials).  In particular, our cross-correlation method works for {\em any} smoothing scale \So; there is no requirement that this scale be large (although, strictly speaking, one does require that $\So < s$, i.e., that the smoothing scale be larger than that used for defining the halos in the first place).  

There are two important lessons here.  First, treating the $\So\to 0$ limit of the bias coefficients as though they are arbitrary is risky: one must be careful to ensure that the implied conditional distribution function is sensible (e.g. positive definite).  Except for the coefficients which come from the more physically motivated excursion set approach, this is rarely ever done.  We return to this point in the next subsection.  The second lesson is that cross-correlating with appropriate transformations of the mass field may be an efficient way of isolating the different large scale bias coefficients from one another.  One view of this second lesson is to contrast it with the usual probe of higher order bias factors:  2-point statistics constrain $b_1$, 3-point statistics constraint both $b_1$ and $b_2$, and so on (Sefusatti \& Scoccimarro 2005; Smith et al. 2007; Pollack \etal\ 2012).  Since the Hermite polynomials here are polynomials, one may think of the transformation as picking out that combination of $n$-point functions which isolates the dependence on $b_n$.  The analysis above suggests that, once the appropriate transformation has been made, $b_n$ can be determined by a real-space cross-correlation measurement alone, and this cross-correlation be made on {\em any} smoothing scale \So; there is no requirement that this scale be large.

\subsection{Scale dependence from appropriate averaging}
\label{pvsq}
The previous subsection noted that a naive averaging of $\avg{1+\delh|\delo}$ over a Gaussian distribution appeared to return the large-scale $\So\to 0$ bias factors.  However, the correct distribution over which to average is not a Gaussian, but 
\be
q_{\rm u}(\delo,\So;\delc) = \frac1{\sqrt{2\pi \So}} \left[ {\rm
    e}^{-\delo^2/2\So} - {\rm e}^{-(2\delc-\delo)^2/2\So} \right] \,,
\label{chandra-q}
\ee
where $\delo < \delc$ (Sheth \& Lemson 1999).  This is because $q_{\rm u}(\delo,\So;\delc)$ gives the probability that the walk had height \delo\ at scale \So, and remained below the barrier \delc\ on all scales $S<\So$ (Chandrasekhar 1943).
It is easy to check that 
\be
f_{\rm u}(s) = \int_{-\infty}^{\delc}\der\delo\, 
                  f_{\rm u}(s|\delo,\So)\,q_{\rm u}(\delo,\So;\delc)\,,
\label{fc-shk-correct}
\ee
as it should.  

For similar reasons, whenever one deals with the conditional mean \avg{1+\delh|\delo}, the appropriate way to compute cross correlations between the halo overdensity field and the mass is by averaging over
$q$ and not $p$, and this generically makes the measured coefficients
depend on scale \So\ as we discuss below.
For $n=1$, this yields 
\begin{align}
 \frac{\delc}{\So}\avg{\delo\avg{1+\delta_h|\delo}}_q &= H_2(\nu)
               + (\nu_{10}^2 + 1)\erfc{\nu_{10}/\sqrt{2}} \notag\\
               &\ph{H_2(\nu) }\qquad - \sqrt{2\nu_{10}^2/\pi}\,e^{-\nu_{10}^2/2}
 \label{b1q}
\end{align}
where 
$\nu_{10}^2 = \nu^2\, (s/\So - 1)$
(equation~17 in Sheth \& Lemson 1999).  
Note that, in contrast to the previous calculation, this quantity
yields $H_2(\nu)$ only in the limit $\So\to 0$.  Similarly, averaging $(1+\delh) H_n$ over $q_{\rm u}$ yields a more complicated function of \So.  

\begin{figure}
 \centering
 \includegraphics[width=\hsize]{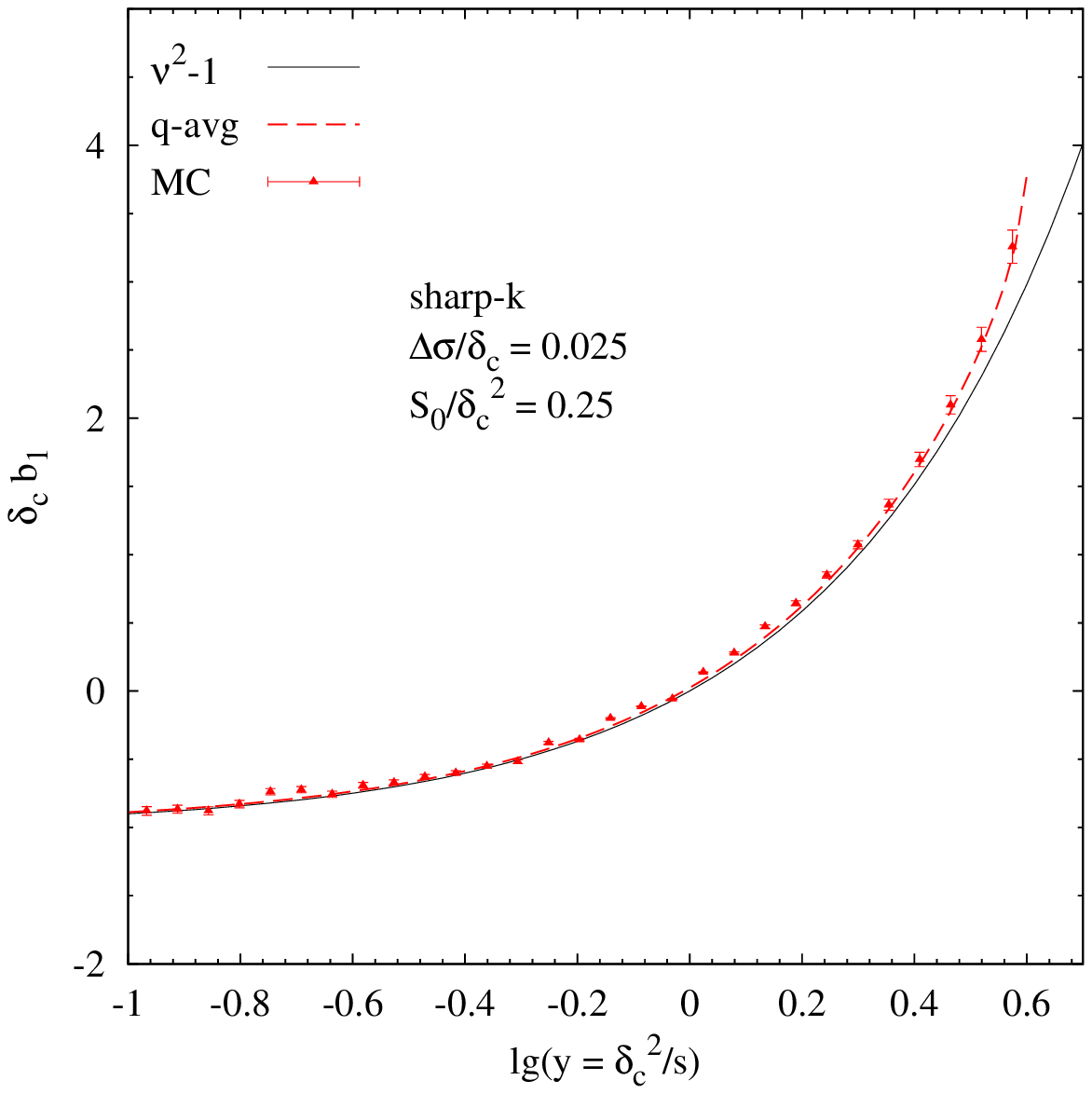}
 \includegraphics[width=\hsize]{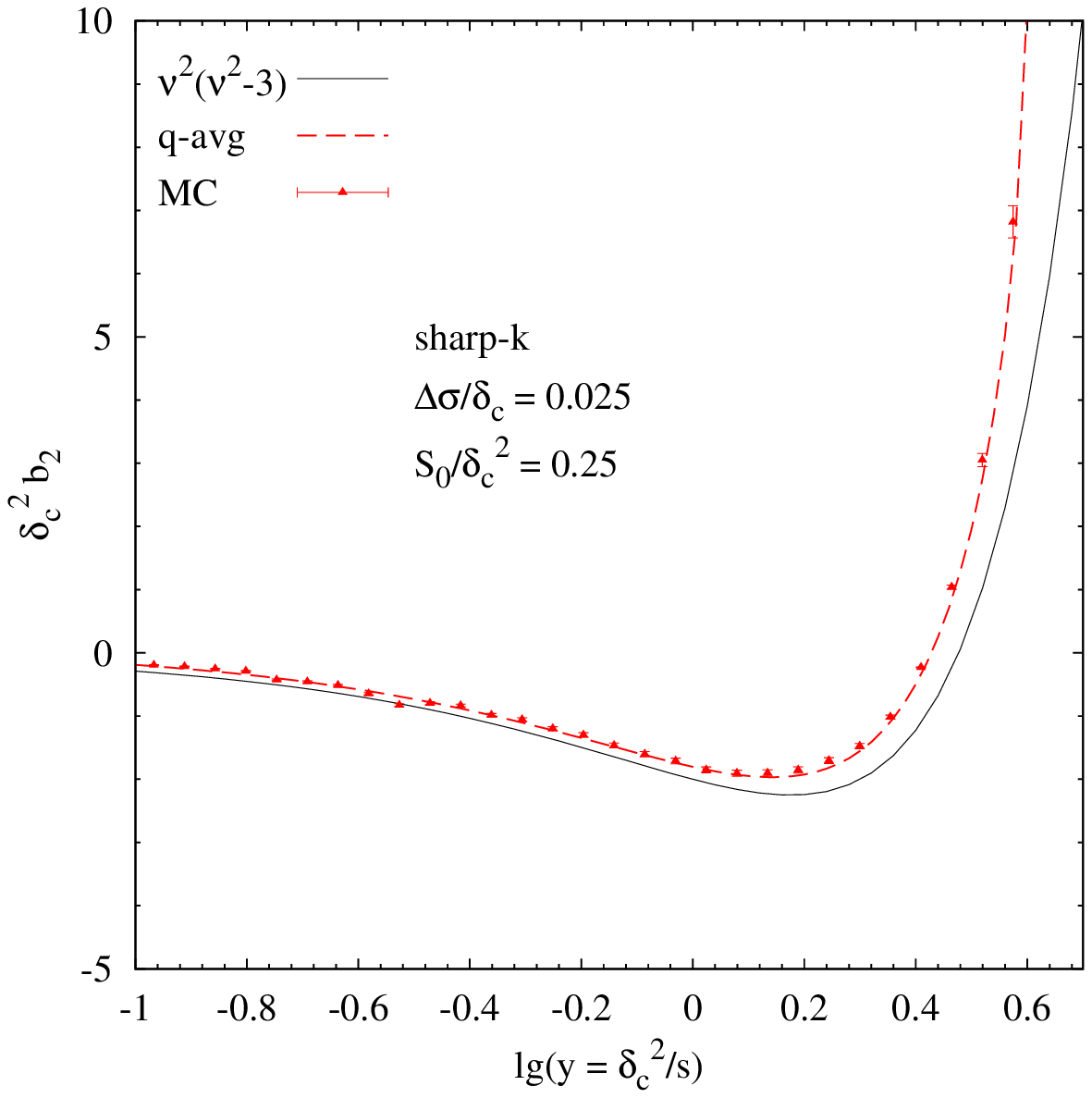}
 \caption{Monte-Carlo measurement of the cross-correlation between 
          $\delh$ and $\delo$ (top) and $H_2$ (bottom), for walks 
          with uncorrelated steps. See the main text for a 
          description of the measurement.
          Solid curves show the prediction 
          associated with averaging over a Gaussian distribution, 
          as is commonly done, and which the main text argued was  
          inappropriate, and dashed curves show the result of 
          averaging using $q$ of equation~(\ref{chandra-q}).  }
 \label{sharpk}
\end{figure}

We have verified these analytical arguments in a comparison with
numerical results.
The symbols in Figure~\ref{sharpk} show a measurement of the
cross-correlation between $\delo$ and $(1+\delh)$ in a Monte Carlo
simulation of random walks with uncorrelated steps. We generate these walks
by accumulating independent Gaussian draws, each with zero mean and
variance $(\Del\sig)^2$. For each such walk, we note the scale $s$ at
which it first crossed a constant barrier \delc, as well as its height \delo\
at a chosen scale \So. The Figure shows results for
$\Del\sig/\delc=0.025$ and $\So/\delc^2=0.25$. 
To measure the correlation in a given bin in $y=\delc^2/s$, we identify those walks that first cross \delc\ in this bin. If these are $N_y$ in number, we compute the mean $\sum_{j=1}^{N_y}H_n(\del_{0j}/\sqrt{\So})/N_y$ where $\del_{0j}$ is the height at \So\ of the $j^{\rm th}$ such walk. Dividing this mean by $\So^{n/2}$ gives the numerical estimate of $b_n$. Since the first-crossing of \delc\ for these walks is at $s>\So$ by construction, this measurement is a $q$-averaged one. 

The two panels show the measurements for $n=1$ and $2$, and the solid and dashed curves show the analytic result of averaging using $p$ and $q$, respectively.  The solid curve remains the same for all
\So\ (equation~\ref{bn-crosscorr-shk}) but the dashed one does not
(e.g. equation~\ref{b1q}).  Therefore, the difference between the
solid and dashed predictions depends on \So; we have checked that
averaging over $q_{\rm u}$ always yields the correct, scale-dependent
value. 

While this agreement demonstrates that we have a good understanding of
just what it is that the excursion set returns, and of the
\So\ scale-dependence of the bias coefficients returned by a
cross-correlation measurement -- it has also shown that averaging
equation~(\ref{fc-LC}) over a Gaussian distribution (rather than
$q_{\rm u}$) will lead to incorrect estimates of the bias factors and
their scale dependence.

The fact that $q\ne p$ leads to measureable differences suggests that
unless one has a good model of how both $f$ {\em and} $q$ depend on
scale, one must use large survey volumes (to ensure one is safely in
the $\So\to0$ limit) if halo bias (e.g. equation~\ref{bias-trad-shk}) is to
constrain parameters.  At the moment, this understanding exists only
for the special case of predictions based on walks with uncorrelated
steps.  Unfortunately, for the $q$-averaging which we have argued is
the more appropriate, it is not straightforward to separate out the
scale independent terms $H_n(\nu)$ from those which depend on
\So\ (e.g., through $\nu_{10}$).  If it were, we would be able to
derive cosmological constraints from smaller volumes.  
As it stands, if walks with uncorrelated steps were a realistic model, then for halos with $\nu\sim1.3$ (mass $m\sim10^{13}h^{-1}\Msun$ or Lagrangian scales $R\sim3h^{-1}$Mpc), in order to achieve percent level accuracy in predicting the scale-independent $b_1$ ($b_2$), one would need to work at scales $\So/\delc^2 \simeq 0.155$ ($0.115$) or Lagrangian scales $R_0 \sim 10h^{-1}$Mpc ($14h^{-1}$Mpc).
We now turn to a study of these issues for the more realistic case of walks with correlated steps.

\section{The excursion set approach with correlated steps}
\label{corrsteps}
We would like to extend the analysis of the previous section to include the effects of correlated steps.  To do so, we must first set up some notation.  

\subsection{Notation}
\label{notation}

Let us recall some standard results regarding Gaussian distributions, which we will use frequently. 
If the joint distribution $p(x_1,x_2)$ for two variables is the bivariate Gaussian with zero mean, then
\begin{align}
  p(x_1,x_2) &= p_{\rm G}(\xb;\C)
  \equiv \frac{{\rm e}^{-\frac12 \xb^{\rm T} \C^{-1} \, \xb}}{
    \sqrt{(2\pi)^2\mathrm{Det}[\C]}}\,,
\end{align}
where $\C$ is the covariance matrix $\C_{ij} = \avg{x_ix_j}$.

If the joint distribution $p(x_1,x_2,x_3)$ for three variables is a trivariate Gaussian, then
the conditional distribution $p(x_1,x_2|x_3)$ is also a bivariate Gaussian:
\begin{align}
p(x_1,x_2|x_3) &= p_{\rm G}(\xb-\bar\xb;\C-\tilde\cb)\,,
\label{condbivarGauss}
\end{align}
where the conditional mean $\bar\xb$ is proportional to $x_3$,
\begin{align}
  \bar\xb &= x_3\bigg(\frac{\avg{x_1x_3}}{\avg{x_3^2}},
  \frac{\avg{x_2x_3}}{\avg{x_3^2}}\bigg)\,.
\label{xbarx3}
\end{align}
The ``correction'' to the covariance matrix $\tilde\cb$ accounts 
for that part of the correlation between $x_1$ and $x_2$ which is due 
to a correlation with $x_3$. Its components are
$\tilde\cb_{11} = \avg{x_1x_3}^2/\avg{x_3^2}$, 
$\tilde\cb_{22} = \avg{x_2x_3}^2/\avg{x_3^2}$ and 
$\tilde\cb_{12} = \avg{x_1x_3}\avg{x_3x_2}/\avg{x_3^2}$.

In the excursion set framework one is interested in $p(\delta,\delta')$
and $p(\del,\del'|\delo)$, where $\del^\prime$ is the ``curvature'' of 
the walk at scale $s$, $\del^\prime = \der\del/\der s$. Since all three quantities $\delta$, $\delta'$ and $\delo$ are essentially linear combinations of the underlying Gaussian-distributed Fourier modes, both these distributions are also Gaussian.
In this case $\avg{\del^2}=s$, $\avg{\delta\delta'}=(1/2)(\der/\der s)\avg{\delta^2} = 1/2$ and 
$\avg{\delo^2}=\So$, and the relevant quantities read
\begin{equation}
  \C = \left[\!
  \begin{array}{cc} s & 1/2 \\ 1/2 & \avg{\!\del'^2\!} \end{array}
  \!\right], \quad
  \tilde\cb = \frac{\Sc^2}{\So^2}\frac{\So}{s}\!
  \left[\!
  \begin{array}{cc} s & \epc/2 \\ \epc/2 & \epc^2/4s \end{array}
  \!\right]
\label{C2}
\end{equation}
and
\begin{equation}
  \bar\xb = \delo \frac{\Sc}{\So} \left(1 , \frac{\epc}{2s} \right)\,,
\label{condmean}
\end{equation}
where 
\begin{equation}
  \Sc \equiv \avg{\del\delo}
  \quad \textrm{and} \quad 
  \epc \equiv 2 s \frac{\avg{\del^\prime\delo}}{\avg{\del\delo}}\,.
  \label{Scross-def}
\end{equation}
For a Gaussian filter, $W(kR)= \exp(-k^2R^2/2)$, one has
$ \Sc = \sig_{0\times}^2$ and 
$\epc=\sig_{1\times}^2\sig_0^2/\sig_{0\times}^2\sig_1^2$, where
\begin{align}
  \sig_{j\times}^2 &= \int\frac{\der k}{k} \frac{k^3P(k)}{2\pi^2}
  k^{2j}W(kR)W(kR_0)\,,
\label{epc-gaussian}
\\
  \sig_j^2 &= \int\frac{\der k}{k} \frac{k^3P(k)}{2\pi^2}
  k^{2j}W^2(kR)\,.
\end{align}
If, in addition, $P(k)\propto k^n$, then 
\begin{align}
\frac{\Sc}{\So} &= 2^{(n+3)/2}\left(1+\left(\So/s\right)^{2/(n+3)}\right)^{-(n+3)/2}\,,
\label{Sc-gausspowlaw}\\
  \epc 
  &=2(\So/s)\left(1+\left(\So/s\right)^{2/(n+3)}\right)^{-1}\,.
\label{epc-gausspowlaw}
\end{align}
We will also use the same notation $p_{\rm G}(z;\sig^2)$ to denote a one-dimensional Gaussian distribution when there is no scope for confusion.

\subsection{The unconditional distribution}
Although our goal is to write down the analogue of equation~(\ref{fc-LC}) for the first crossing distribution associated with walks which are conditioned to pass through \delo\ on scale \So, our first step is to write down the unconstrained distribution. As shown by Musso \& Sheth (2012), for a constant barrier of height \delc\ the latter is well-approximated by
\be
 f(s) = \int_0^\infty \der \del^\prime \del^\prime p(\delc,\del^\prime)\,, 
\label{fc-MS}
\ee
where 
$p(\delc,\del^\prime)$ is the bivariate Gaussian
$ p_{\rm G}(\delc,\del^\prime;\C)$ with covariance matrix 
given in \eqn{C2}.

The integral in \eqn{fc-MS} can be performed analytically and leads to
\begin{equation}
sf(s) = \frac{\nu{\rm e}^{-\nu^2/2}}{2\,\sqrt{2\pi}}\bigg[ 
  \frac{1+\erf{\Gamma\nu/\sqrt{2}}}{2} +
  \frac{{\rm e}^{-\Gamma^2\nu^2/2}}{\sqrt{2\pi}\Gamma\nu} \bigg],
\label{sfsMS12}
\end{equation}
with
\be
 \Gamma^2\equiv\frac{\gamma^2}{(1-\gamma^2)} \quad {\rm and}\quad
 \gamma^2\equiv
  \frac{\avg{\del\del^\prime\!}^2}{\avg{\del^2}\!\avg{\del^{\prime2}\!}}\,.
\label{G2}
\ee
(Equation~\ref{sfsMS12} corrects a typo in equation~6 of the published version of Musso \& Sheth 2012.)
For later convenience, the same can also be written as
\begin{equation}
  sf(s) = \frac{{\rm e}^{-\nu^2(1+\Gamma^2)/2}}{(2\pi)(2\Gamma)} 
  \left(1 + A\right)\,, 
\label{fc-MS-explicit}
\end{equation}
where 
\be
A \equiv A(\nu) = \frac12 \left[1+\erf{\frac{\Gamma\nu}{\sqrt{2}}} \right] 
 \sqrt{2\pi}\,\Gamma\nu \,{\rm e}^{\Gamma^2\nu^2/2}\,.
\label{A-def}
\ee
Musso \& Sheth (2012) showed that this approximation (as well as its generalisation to moving barriers) works extremely well over a large range of scales for a range of choices of power spectra and filters (including TopHat filtered LCDM) when compared with Monte Carlo solutions of the first crossing problem.  (It breaks down in the limit in which the walks must have taken many steps to cross the barrier.)  Our final analytic results in this paper will be valid for arbitrary power spectra and filters for a constant barrier. However, since explicit expressions for various quantities greatly simplify for the choice of Gaussian smoothing of power law power spectra, we will show comparisons with Monte Carlo solutions for the latter.
For Gaussian smoothing, $\gamma=\sig_1^2/\sig_0\sig_2$.

\begin{figure}
 \centering
 \includegraphics[width=\hsize]{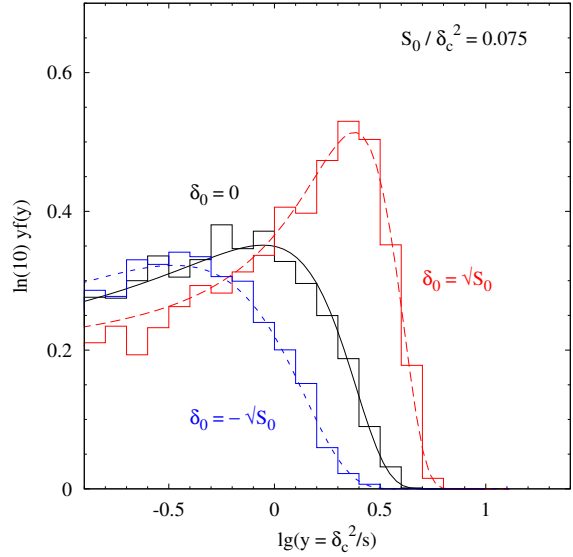}
 \includegraphics[width=\hsize]{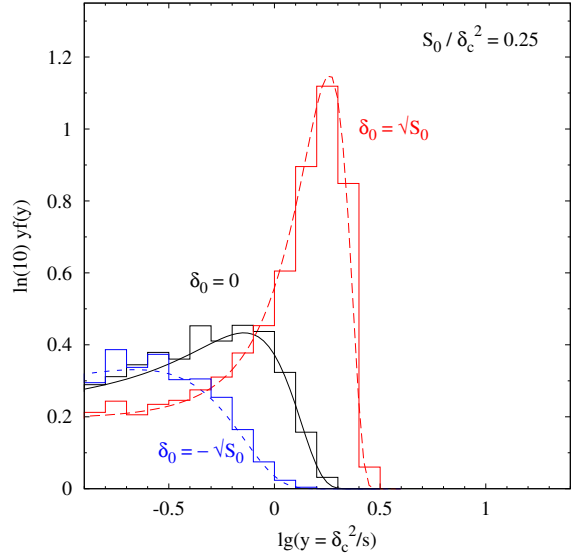}
 \caption{First crossing distribution of a barrier of height 
          $\delta_{\rm c}$ by the subset of walks which 
          are conditioned to pass through $(\delta_0,S_0)$, for 
          a few choices of $\delta_0$ (as labelled). 
          Short dashed, solid and long-dashed curves show the 
          analytic prediction from \eqn{condfc-explicit}
          for Gaussian smoothing of a Gaussian field with 
          $P(k)\propto k^{-1.2}$.}
 \label{vfvCond}
\end{figure}
\subsection{The conditional distribution}
\label{condfc-analytical}
Musso \& Sheth (2012) argued that the conditional distribution corresponding to \eqref{fc-MS} is simply
\be
f(s|\delo,\So) = \int_0^\infty \der \del^\prime \del^\prime
p(\delc,\del^\prime|\delo)\,,
\label{fc-MS-cond}
\ee
where $p(\delc,\del^\prime|\delo)$ is the probability that the walk had a height \delc\ and curvature $\del^\prime$ at scale $s$, given that it passed through \delo\ at scale $\So<s$. In principle, one is really interested in imposing the stronger condition that the walk must have passed through $(\delo,\So)$ without having crossed \delc\ before \So. We will return to this point later and argue that the effects of ignoring this stronger requirement are small.

The conditional distribution $p(\delc,\del^\prime|\delo)$ is the bivariate Gaussian 
\be
 p(\delc,\del^\prime|\delo) = p_{\rm G}(\Del-\bar\xb;\C-\tilde\cb)\,, 
 \label{condgauss}
\ee
with $\Del\equiv(\delc,\del')$, \C\ and $\tilde\cb$ given by \eqn{C2} 
and the conditional mean $\bar\xb$ given by \eqn{condmean}.
For generic power spectra and filters, the integral in \eqn{fc-MS-cond} can be performed analytically, exactly as in the case of \eqn{fc-MS}, and expressed in terms of $\Sc/\So$ and \epc. The result is
\begin{align}
f(s|\delo,\So) &= \frac{\dprbar\,{\rm e}^{-\dcr^2/2sQ}}{\sqrt{2\pi s Q}}\notag\\
  &\times\left[\frac{1+\erf{\dprbar/\sqrt{2}\sigdpr}}{2}
  + \frac{{\rm e}^{-\dprbarsq/2\sigdprsq}}{\sqrt{2\pi}(\dprbar/\sigdpr)}
  \right]\,,
\label{condfc-explicit}
\end{align}
where 
\be
 \dcr \equiv \delc - \delo\frac{\Sc}{\So} ~~;~~ Q\equiv 1-\left(\frac{\Sc}{\So}\right)^2\frac{\So}{s}\,,
\label{dcr-Q}
\ee
\be
 \dprbar \equiv \avg{\!\del^\prime|\delc,\delo\!}
 = \frac1{2sQ}\!\left[ \dcr +
 \epc\frac{\Sc}{\So}\bigg(\delo - \delc\frac{\Sc}{\So}\frac{\So}{s}\bigg)\right]\,,
 \label{dprbar}
\ee
and 
\be
 \sigdprsq \equiv {\rm Var}(\del^\prime|\delc,\delo)
 = \frac1{4\Gamma^2s}\! \left[1-\frac{\Gamma^2\So}{Qs} 
 \frac{\Sc^2 (1-\epc)^2}{\So^2} \right]\,.
\label{sigdprsq}
\ee
Note that, in contrast to equation~(\ref{fc-LC}), this expression for the conditional distribution remains positive definite even when $\delo > \delc$, although it is understood that only $\delo \le \delc$ is sensible.  For future reference, the sharp $k$-space filter has $\Sc/\So = 1$ and $\epc=0$; its conditional crossing distribution, \eqn{fc-MS-explicit}, corresponds to inserting these values in \eqn{condfc-explicit} and replacing the term in square brackets with a factor of 2.

\subsection{Comparison with Monte Carlo solution}
\label{condfc-MC}

\fig{vfvCond} compares the prediction in \eqn{condfc-explicit} with a Monte Carlo solution of the conditional first crossing distribution. The comparison is for Gaussian filtered random walks using a power spectrum $P(k)\propto k^{-1.2}$. The numerical treatment uses the algorithm of Bond \etal\ (1991) and was described in Paranjape \etal\ (2012). The histograms are the same as in Figure 6 of Paranjape \etal\ and show the distribution of first crossing scales for a constant barrier, for walks that were required to pass through the indicated values of \delo\ at scale \So, for two choices of \So. 
We see that the analytic prediction works very well in describing the numerical solution.

\begin{figure}
 \centering
 \includegraphics[width=\hsize]{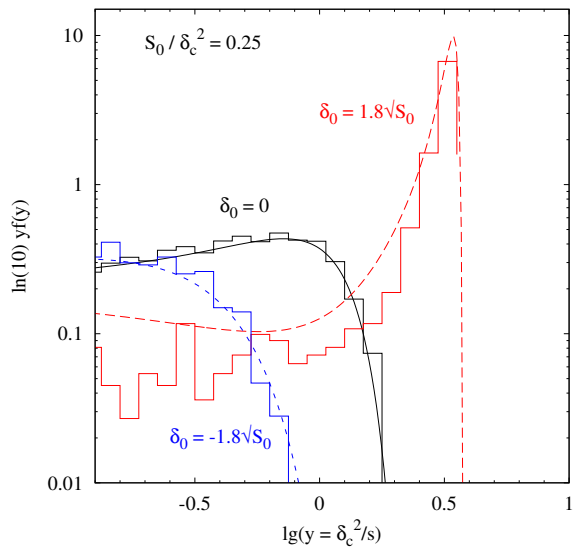}
 \caption{
   Same as lower panel of \fig{vfvCond}, but for a larger value 
   of $|\delo|$, and note that now the $y$-axis is on a log scale.  
   The analytic prediction \eqn{condfc-explicit} 
   for $\delo=0.9\delc$ describes the sharp peak in the numerical 
   solution remarkably accurately. 
   It begins overestimating the numerical solution around 
   ${\rm log}_{10}\nu^2 \simeq 0.4$, at which point about 75\% 
   of the probability has been accounted for (see text for why 
   this happens).  
 }
 \label{vfvCond-2}
\end{figure}

This good agreement is despite the fact that \eqn{fc-MS-cond} formally ignores walks which might have crossed the barrier prior to \So. This can be understood by the fact that the values of \delo\ being considered in \fig{vfvCond} are significantly smaller than \delc, so that very few of the walks would have reached the barrier prior to \So\ and then returned to pass through \delo\ at \So. One can then ask whether the expression in \eqn{condfc-explicit} would continue to be accurate even for $\delo\lesssim\delc$, since this is the regime of interest for calculations of merger rates. 

We test this in \fig{vfvCond-2}, which compares \eqn{condfc-explicit} with the Monte Carlo solution for the same choice of conditioning scale \So\ as in the lower panel of \fig{vfvCond}, but with a larger magnitude for \delo\ which is now $|\delo|=0.9\delc$. We see that for $\delo=+0.9\delc$, the numerical solution has a sharp peak which is very well described by \eqn{condfc-explicit}. The latter starts overestimating the numerical answer around ${\rm log}_{10}\nu^2 \simeq 0.4$, which can be understood as follows. 

Paranjape \etal\ (2012) demonstrated in their Figure 7 that the numerical conditional distributions are, to a good approximation, related to the corresponding unconditional one by a simple scaling relation which sends $\nu\to\nu_{10}=\dcr/\sqrt{sQ}$ in the unconditional distribution. This is also approximately true of the analytic expression in \eqn{condfc-explicit}. Since \eqn{fc-MS-explicit} is not a good approximation to the \emph{unconditional} first crossing distribution at small values of $\nu$ (Musso \& Sheth 2012), it follows that the corresponding analytic \emph{conditional} distribution will not be a good approximation at small $\nu_{10}$. One can check that, for the choices of \So\ and \delo\ in \fig{vfvCond-2}, $\nu_{10}$ actually passes through zero and becomes negative around ${\rm log}_{10}\nu^2 \simeq 0.5$. So the mismatch between the analytic prediction and the numerical solution is not surprising.  In practice, $\int_{0.4\ln 10}^{-\ln S_0} d\ln y\, yf(y|S_0) = 0.75$, indicating that the prediction 
is inaccurate only for the 25\% which cross at the largest values of $s$ (smallest values of $y$).  

\subsection{Halo bias with correlated steps}
\label{bias}
Now that we have in hand a good approximation to the conditional first crossing distribution, we can turn to the associated description of halo bias. 

\begin{figure}
 \centering
 \includegraphics[width=\hsize]{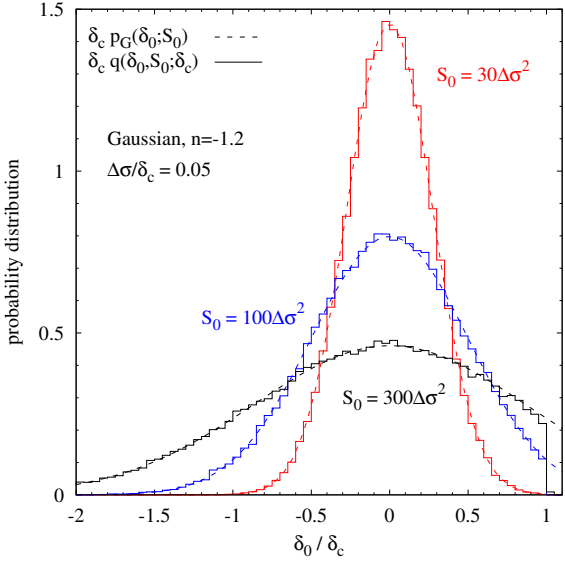}
 \caption{Distribution of the height $\delta_0$ on scale \So\, 
   of walks which have not crossed \delc\ prior to \So, for a range 
   of choices of \So\ (histograms), measured in the same Monte-Carlo 
   simulations which were used to make the \figs{vfvCond} and~\ref{vfvCond-2}.  
   Dashed curves show that a Gaussian, truncated at $\delo=\delc$, 
   provides a good approximation.  
   Note that $\So/\delta_c^2 = 300\times 0.05^2 = 3/4$ corresponds 
   to smoothing scales which are of order that associated with a 
   typical halo:  therefore, if one restricts attention to smaller 
   \So, then ignoring the truncation of the Gaussian should be a 
   good approximation.}  
 \label{correlated-q}
\end{figure}

The first issue that we would like to address is if Hermite polynomials of the smoothed matter density field are still special.  
Appendix \ref{app-matsubara} suggests that they are, as long as the underlying matter density field is Gaussian. More formally, we show there that the r\^ole of the Hermite polynomial $H_n(\delo/\sqrt{\So})$ in the average is that of removing from it all the disconnected parts, so that only the connected part of the expectation value of $\delo^n$ remains.

Secondly, for reasons discussed in section~\ref{pvsq}, in principle we must specify the probability distribution $q(\delo,\So;\delc)$ to be used in the average. In the present case, $q$ is not known analytically.  However, in the spirit of Musso \& Sheth (2012), we can argue that the error in ignoring the difference between $p$ and $q$ is of the same order as that already included in $f(s|\delo,\So)$. Indeed, the fact that the conditional distributions shown in \fig{vfvCond} are such an accurate description of the numerical solution means that, in this case, the approximation is consistent.  This is a consequence of the fact that for correlated steps zig-zags are exponentially rare at small $S_0$; in this limit, $p\approx q$. 
We can test this explicitly by looking directly at the distribution of $q$ in our Monte-Carlos.  \fig{correlated-q} shows that, for \So\ values which are smaller than those associated with typical halos, the difference between $q$ and the Gaussian is almost negligible. 

Motivated by this simplification, let us \emph{define} the real-space
bias coefficients associated with the conditional distribution
$f(s|\delo,\So)$ using
\begin{align}
b_n &\equiv \frac1{\So^{n/2}} \avg{(1+\delh) H_n(\delo/\sqrt{\So})} \notag\\
&= \int_{-\infty}^\infty\der\delo\,p_{\rm G}(\delo;\So) \avg{1+\delh|\delo,\So} H_n(\delo/\sqrt{\So})\,,
\label{bn-crosscorr}
\end{align}
with \avg{1+\delh|\delo,\So}\ given in \eqn{delhgivendel0}. 
Below we will show comparisons between numerical measurements  of
these quantities ($q$-averaged by construction) with analytic results
using the $p$-averaged expression in the second line of
\eqref{bn-crosscorr}. From the discussion above, we expect these to
match well at least for the smallest \So\ shown in
\fig{correlated-q}. 

For $f(s|\delo,\So)$ given by \eqn{fc-MS-cond}, some algebra brings these into the form (see Appendix~\ref{app-biascorr-def})
\be
b_n = \frac{(-\Sc/\So)^n}{f(s)} \int_0^\infty\der\del^\prime\del^\prime 
\left( \frac{\p}{\p\delc} +
\frac{\epc}{2s}\frac{\p}{\p\del^\prime} \right)^{\!n} 
\!p(\delc,\del^\prime)\,,
\label{bias-corr}
\ee
with $f(s)$ given in \eqn{fc-MS}. 
Appendix~\ref{app-condfc-taylor} shows that 
\begin{align}
f(s|\delo,\So) &= \sum_{n=0}^\infty\frac{\delo^n}{n!}
 \left(-\frac{\Sc}{\So}\right)^n 
\int_0^\infty\der\del^\prime\del^\prime \notag\\
&\ph{\sum}
\times\left( \frac{\p}{\p\delc} +
\frac{\epc}{2s}\frac{\p}{\p\del^\prime} \right)^{\!n}\! 
p_{\rm G}(\Del;\C-\tilde\cb)\,,
\label{fc-MS-cond-Taylor}
\end{align}
holds exactly for the distribution \eqref{fc-MS-cond}, where $\Del=(\delc,\del^\prime)$ and the matrices \C\ and $\tilde\cb$ were defined in \eqn{C2}. Since the bivariate Gaussian $p_{\rm G}(\Del;\C)$ is precisely the distribution $p(\delc,\del^\prime)$ that appears in \eqn{bias-corr}, we clearly have
\be
\frac{f(s|\delo,\tilde\cb=0)}{f(s)} = 1 + \sum_{n=1}^\infty\frac{\delo^n}{n!}b_n \,.
\label{bias-corr-full}
\ee
Setting $\tilde\cb=0$ corresponds to the following assignments in \eqn{condfc-explicit}:
\begin{align}
  &Q\to1 ~~; \qquad \sigdpr\sqrt{s}\to(2\Gamma)^{-1}\,;\notag\\
  &\dprbar/\sigdpr \to \Gamma\nu \left[1
  - (\delo/\delc)(\Sc/\So)(1-\epc)\right].
\label{ctilto0}
\end{align}
As a result (see Appendix~\ref{app-bias-explicit}) the bias coefficients can be reduced to:
\be
\delc^nb_n = \left(\frac{\Sc}{\So}\right)^n
 (\alpha_n + \beta_n + \gamma_n)\,,
\label{bn-corr-explicit}
\ee
where
\begin{align}
\alpha_n &=  \nu^nH_n(\nu)\,, ~~ n\geq1\,,
\label{alpha_n}\\
\beta_n &= \frac1{1+A}%
\begin{cases}
-A\left(1-\epc\right)&,\, n=1\\
\left(1-\epc\right)^n(\Gamma\nu)^nH_{n-2}(\Gamma\nu)&,\, n\geq2
\end{cases}
\label{beta_n}\\
\gamma_n &=%
\begin{cases}
0 &,~~ n=1\\
\sum_{k=1}^{n-1}\binom{n}{k}\alpha_k\beta_{n-k} &,~~ n\geq2\,,
\end{cases}
\label{gamma_n}
\end{align}
where $A$ was defined in \eqn{A-def}.  

\begin{figure}
 \centering
 \includegraphics[width=\hsize]{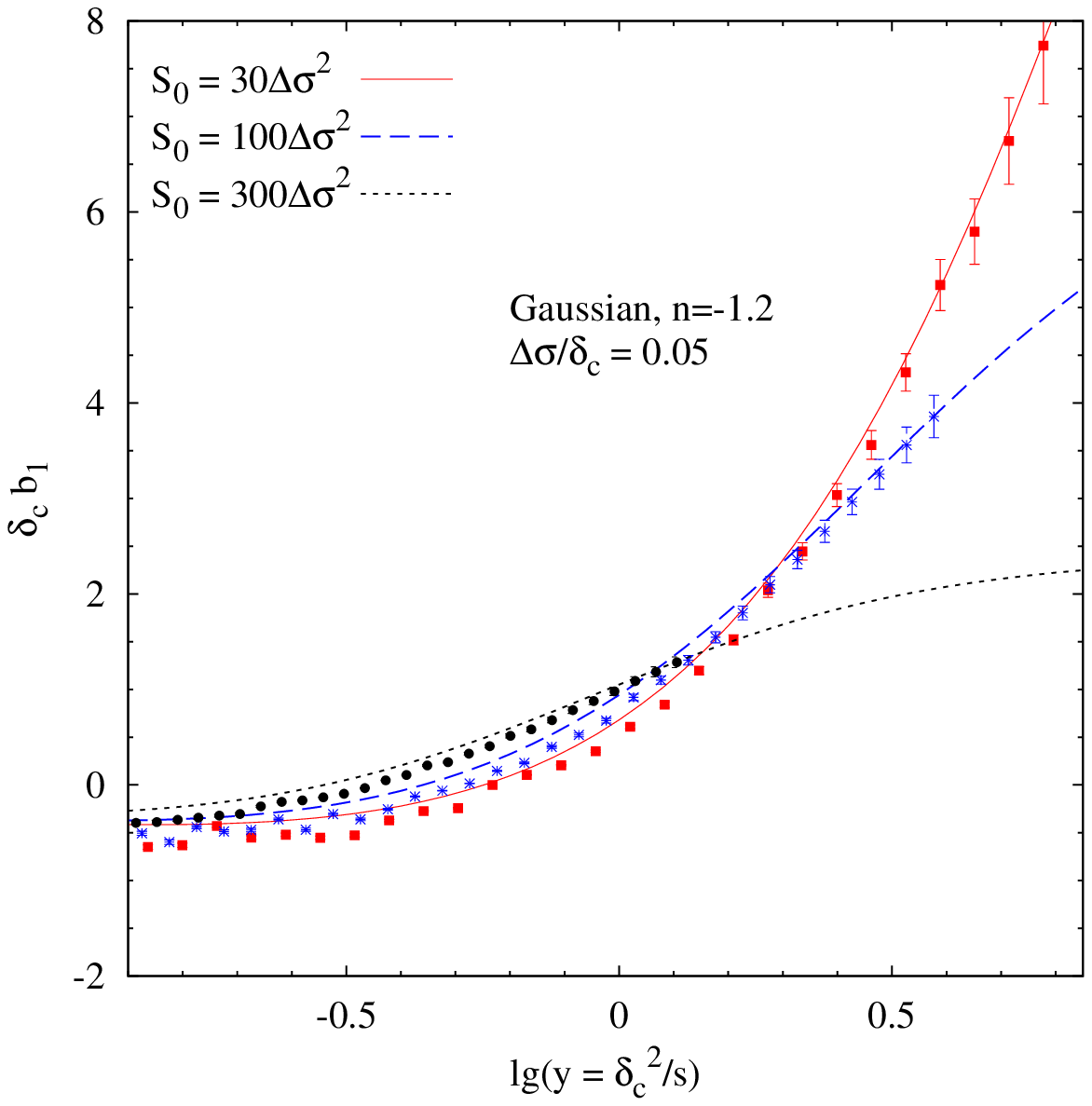}
 \includegraphics[width=\hsize]{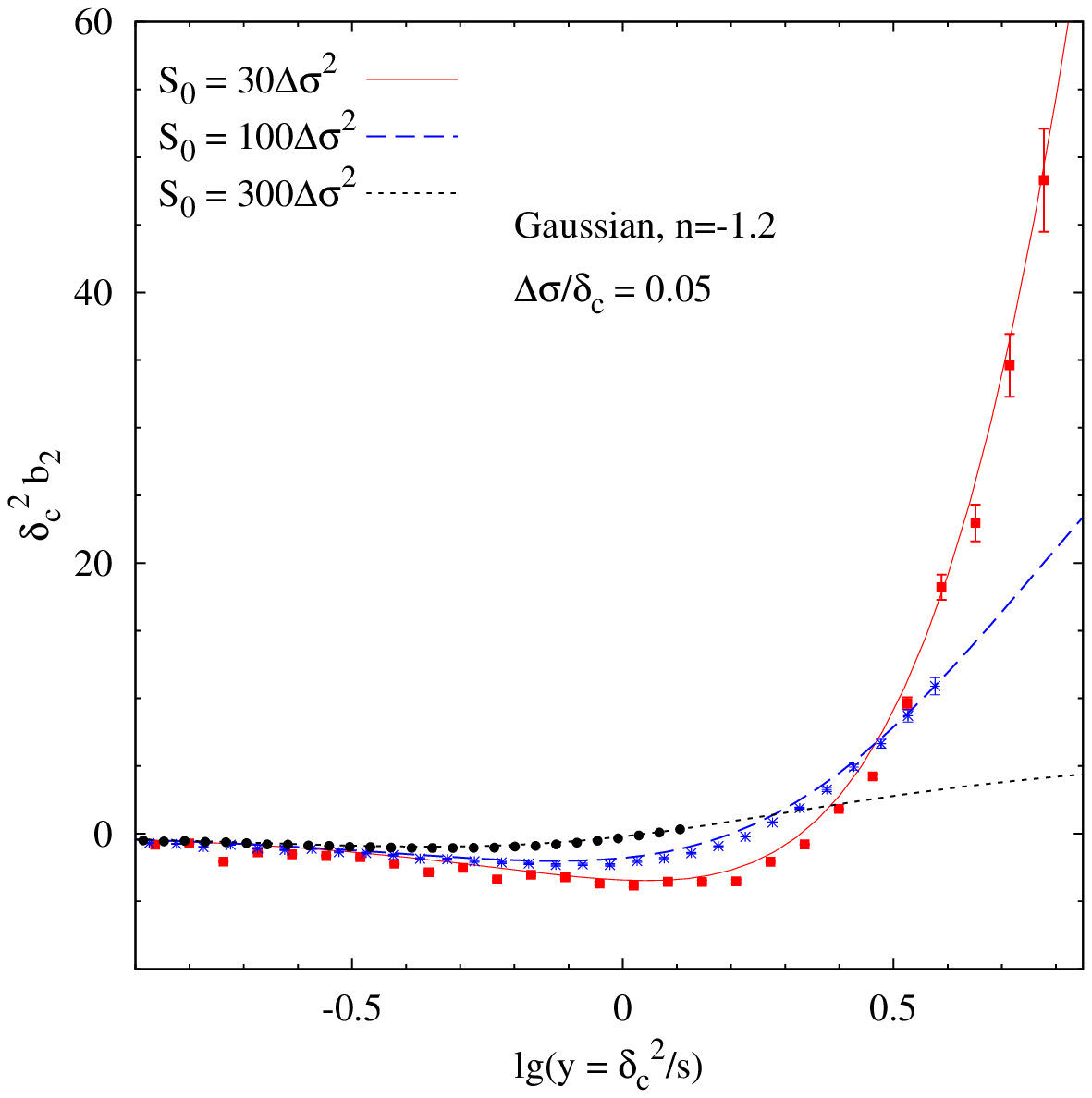}
 \caption{Comparison of bias coefficients $b_1$ and $b_2$ of 
   \eqn{bn-corr-explicit} (smooth curves) with corresponding 
   measurements (points with Poisson errors) in the same Monte Carlo 
   simulations used in \figs{vfvCond} and~\ref{vfvCond-2}, 
   for a range of \So\ values. 
   The measurements were performed as described in section~\ref{pvsq}.
   The analytic prediction clearly tracks both the $s$- and 
   \So-dependence fairly accurately. There are small systematic
   differences, especially at large $s$, which arise because  
   $q\neq p$ at large \So, and our analytic approximation to 
   $f(s|\delo,\So)$ stops being a good approximation when $s\gg\So$. }
 \label{biascompare}
\end{figure}

There are some interesting parallels with the calculation for sharp-$k$ walks, and some important differences. There is obviously a close analogy between the $\So\to 0$ limit of sharp-$k$ walks and the $\tilde\cb\to0$ limit for correlated steps, especially since the matrix $\tilde\cb$ is proportional to \So\ (c.f. equation~\ref{C2}, noting that the factor $\Sc/\So$ becomes constant as $\So\to0$). However, in the present case one is \emph{not} throwing away all the dependence on \So, since factors of \epc\ explicitly appear in the expression for the $b_n$.  In particular, these factors of \epc\ would not have appeared if we had simply taken derivatives of the unconditional distribution (equation~\ref{fc-MS-explicit}) with respect to \delc.  This has an important consequence:  for sharp-$k$ filtering, the quantities $b_n$ were independent of \So, whereas here they depend explicitly on \So.  If we write \eqn{bn-corr-explicit} as 
\be
 b_n\sim (\Sc/\So)^n \sum_{k=0}^n b_{nk}\,\epc^k,
 \label{bnk}
\ee
then it is the quantities $b_{nk}$ (rather than $b_n$) which are scale-independent.  This will be important below when interpreting our results in terms of Fourier-space bias.  Note that the $b_{n0}$ are the peak-background split parameters $f^{-1}(-\p/\p\delc)^nf$ which are of most interest in cosmological applications. This is obvious upon setting $\epc\to0$ in \eqn{bias-corr}.

Since $p\approx q$, in contrast to when steps are uncorrelated, one might expect equation~(\ref{bias-corr}) to be quite accurate.  We test this explicitly in \fig{biascompare} by comparing the results of evaluating the r.h.s. of \eqn{bias-corr} for $n=1$ and $n=2$ with corresponding measurements (performed as described in section~\ref{pvsq}) using the same Monte Carlo simulations that were used in \figs{vfvCond} and~\ref{vfvCond-2}.  
By construction, the numerically estimated quantity is $q$-averaged,
whereas the analytic curves show the Gaussian-averaged coefficients in
\eqn{bn-crosscorr}. The analytic predictions closely track the
measurements over a range of $s$-values for several choices of
\So. There are small systematic deviations which are likely due to a
combination of the facts that $q\neq p$ at large \So\ and that
the analytic prediction fails to be a good approximation at
large $s$.  

Since ignoring the difference between $p$ and $q$ is a good approximation, one might wonder if the effect of \epc\ can also be ignored; naively one expects the $q$-averaging to be irrelevant at small $\So/s$ where \epc\ is also likely to be small. \fig{biasnoep} shows the results for $b_1$ and $b_2$ for one of the choices of \So\ from \fig{biascompare}, comparing the same measurements as in that figure with analytic expressions in which \epc\ is retained as per \eqns{beta_n} and~\eqref{gamma_n} (solid curves) or set to zero by hand (dashed curves). We see that the terms involving \epc\ contribute significantly and must be retained to get an accurate description of the bias.

\subsection{Recovery of scale-independent bias factors}\label{recover}
The bias coefficients in \fig{biascompare} show a strong dependence on the scale \So.  This is rather different from the case of sharp-$k$ filtering, for which the $b_n$ recovered from $p$-averaging (equation~\ref{bn-crosscorr-shk}) were independent of \So.  Indeed, the scale-independence of the recovered $b_n$ was one of our motivations for cross-correlating with the Hermite-transformed field in the first place, so it is interesting to ask if the dependence on \So\ can be removed.

\begin{figure}
 \centering
 \includegraphics[width=0.975\hsize]{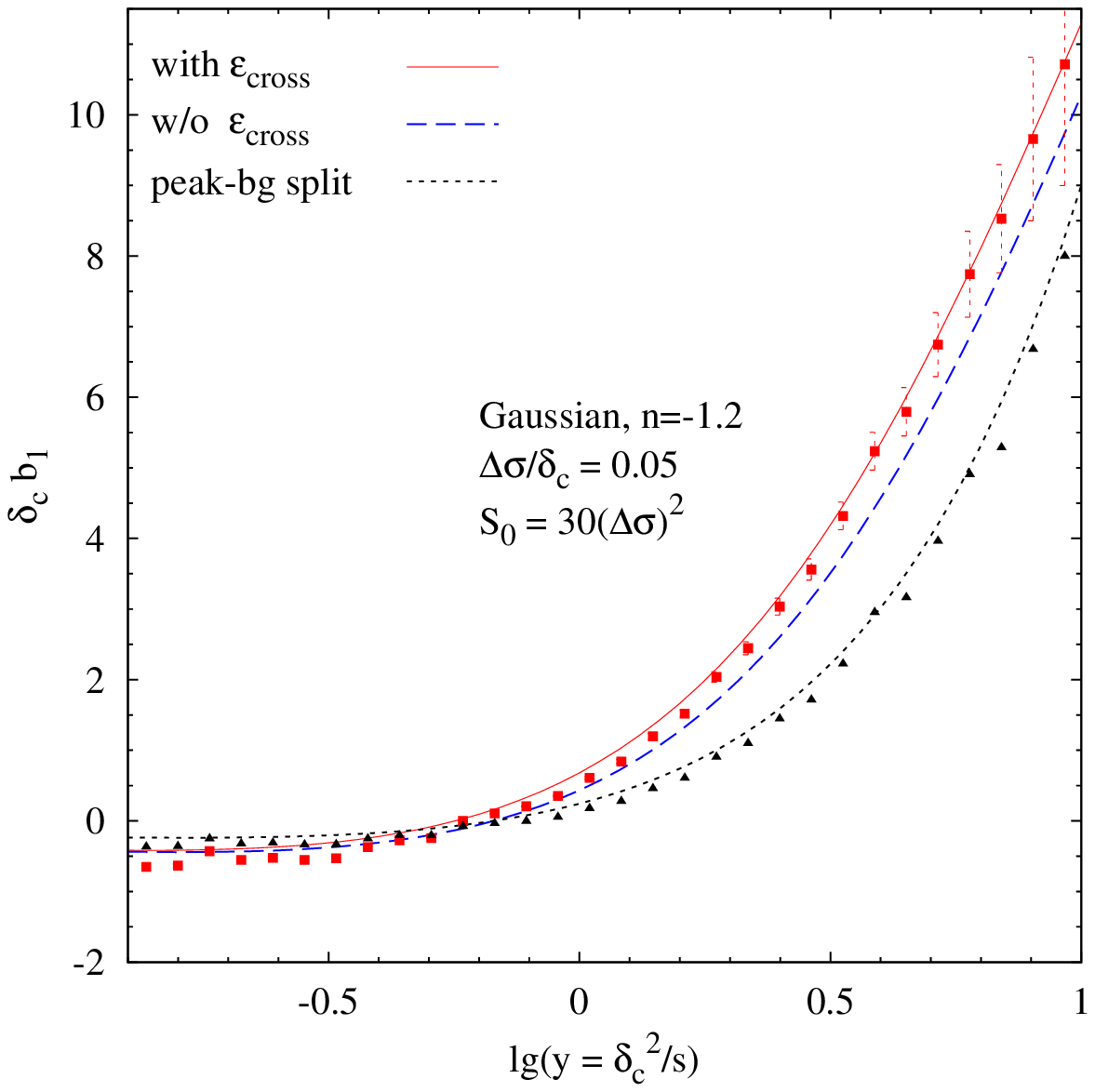}
 \includegraphics[width=0.975\hsize]{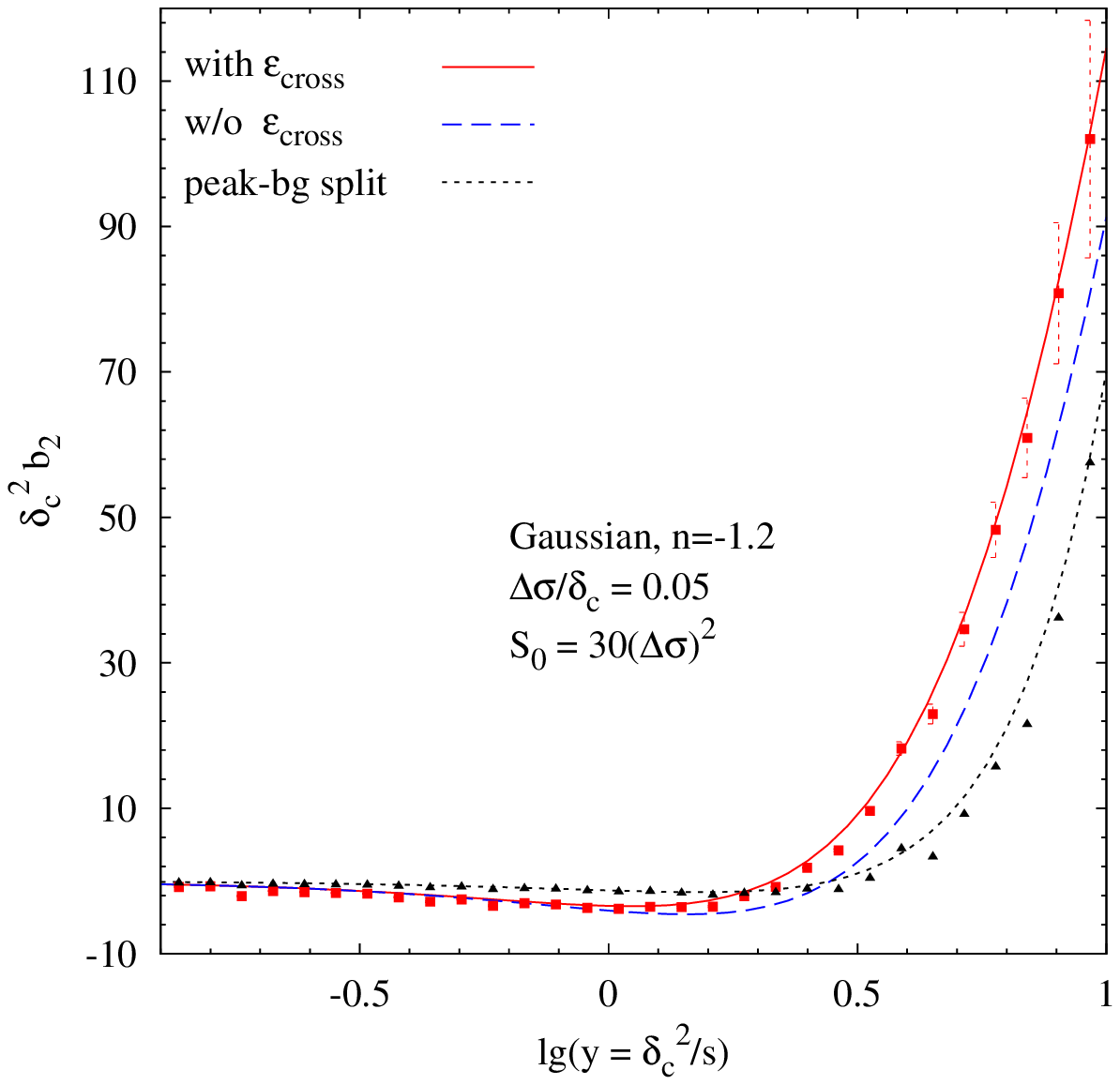}
 \caption{Same as \fig{biascompare} for one choice of \So, but comparing 
   the numerical answer for $b_1$ and $b_2$ with the analytic prediction 
   \eqn{bn-corr-explicit} when the dependence on \epc\ is retained
   (solid red) or set to zero by hand (dashed blue). Clearly,
   retaining the dependence on \epc\ is important, indicating that 
   our method is sensitive to the $k$-dependence of bias (see text).  
   Black triangles show the result of implementing the recursive 
   procedure described in the text for reconstructing the usual 
   ($k$-independent) peak-background split parameters $b_{10}$ and 
   $b_{20}$ (dotted curves) from these measurements.
   Although defined at finite \So, the procedure works well
   in reproducing the \So-independent $b_{n0}$.}
 \label{biasnoep}
\end{figure}

This turns out to be possible because of the following. First, the scale dependence of $b_n$ is almost entirely due to the factors of \Sc\ and \epc\ (the other effect comes from the small difference between $p$ and $q$ averaging). And secondly, \eqn{bn-corr-explicit} shows that the scale-independent $b_{nk}$ are \emph{linearly} related to each other in such a way that measuring $b_1,\ldots,b_n$ is sufficient to recover all the $b_{1k},\ldots b_{nk}$. 

We demonstrate this explicitly for $n=1$ and $2$. For $n=1$, we can write
\begin{align}
 b_1 &= \frac1{\delc} \frac{\Sc}{\So}
  \left[\left(\nu^2 - \frac{A}{1+A}\right) + \epc \frac{A}{1+A}\right] \notag\\
     &\equiv \frac{\Sc}{\So}(b_{10} + \epc b_{11})\,.
\label{b1corr}
\end{align}
Since
\be
 \delc\, b_{11} = \nu^2 - \delc b_{10}\,,
 \label{b10-b11-corr}
\ee
we can estimate 
\be
 \delc\, b_{10} = \frac{\delc\, (\So/\Sc)b_1 - \epc \nu^2}{1 - \epc}.
 \label{b10-est}
\ee
Similarly, 
\be
 b_2 = \left(\frac{\Sc}{\So}\right)^2
       \left(b_{20} + 2\epc\,b_{21} + \epc^2b_{22}\right)\,,
\label{b2corr}
\ee
where the excursion set predictions for the coefficients $b_{2j}$ can be
read off from \eqn{bn-corr-explicit}. For example,
 $\delc^2\,b_{21} = \nu^2(A-\Gamma^2)/(1+A)$.
But, more relevant to the present discussion, 
we find
\begin{align}
 \delc^2b_{21} &= \nu^2(\delc b_{10}-1) - \delc^2b_{20}\notag\\
 \delc^2b_{22} &= \delc^2b_{20} + \nu^2(\nu^2 - 2\delc b_{10}+1)\,.
 \label{b20-etc-relation}
\end{align}
Hence, 
\begin{align}
 \delc^2\,b_{20} = &\frac{1}{\left(1 - \epc\right)^2}\bigg[
  \delc^2 \bigg(\frac{\So}{\Sc}\bigg)^{\!2}b_2  \notag\\
  &- \epc\nu^2 \,\bigg(2\delc \frac{\So}{\Sc}b_1 
     - \epc(\nu^2-1) - 2\bigg)\bigg]\,.
 \label{b20-est}
\end{align}
We have deliberately isolated the peak-background split parameters $b_{n0}$ above. 
From the structure of the coefficients in \eqn{bn-corr-explicit} it is clear that this reconstruction can be extended to the higher order coefficients as well. 

The dotted curves in \fig{biasnoep} show the analytic predictions for
$b_{10}$ and $b_{20}$ from \eqn{bn-corr-explicit}, while the
triangular symbols show the numerical estimates using
\eqns{b10-est},~\eqref{b20-est} and the corresponding measurements of
$b_1$ and $b_2$. Clearly, the reconstruction works well. Moreover,
since we are working at finite \So, our procedure has allowed a simple
and direct estimate of the peak-background split parameters $b_{n0}$
from a measurement of scale-dependent bias, \emph{without} having to
access very large scales.  E.g., the Figure shows results for $\So =
0.075\,\delc^2$, which corresponds to the scale associated with a
$\nu\approx 3.7$ halo and a Lagrangian length scale of $R_0\sim17h^{-1}$Mpc; most other analyses of halo bias are restricted
to length scales which are several times larger.

Another way to see this is to notice that, in the expressions above, $b_n\to b_{n0}$ when $\epc\to 0$.  Since $\epc\to 0$ on large scales, the analysis above shows explicitly that our method for reconstructing $b_{n0}$ works even on the smaller scales where $\epc\ne 0$.  Indeed, although we have concentrated on isolating $b_{n0}$, the analysis above shows that we can isolate the other $b_{nk}$ as well.  For example, having measured $b_1$ and $b_2$ using our Hermite-weighting scheme, and having used equations~(\ref{b10-est}) and~(\ref{b20-est}) to estimate $b_{10}$ and $b_{20}$, \eqn{b20-etc-relation} furnishes estimates of $b_{21}$ and $b_{22}$.  

The expressions above show that our method will break if $\epc=1$, 
which happens when $s\to \So$.  This is not surprising since this is the limit in which the large scale environment is the same as that on which the halo was defined, so our expressions for the conditional distribution are becoming ill-defined.  Since this regime is substantially smaller than the one of most interest in cosmology, we conclude that our method allows a substantial range of interesting scales to provide estimates of the bias factors $b_{nk}$.

\subsection{Real and Fourier-space bias}
\label{biasfourier}
The appearance of \epc\ in the real-space expressions for $b_n$ generically indicates that the bias in Fourier space must be $k$-dependent. This is most easily seen with $b_1$ using a Gaussian filter $W(kR)={\rm e}^{-k^2R^2/2}$. 

Suppose that 
\be
 \delo(\kb)=\del(\kb)W(kR_0)
\ee
and 
\be 
 \delh(\kb) = b_1(\kb) \del(\kb) W(kR)\,,
\ee
so that in real space $\avg{\delh|\delo}=\delo\avg{\delh\delo}/\So$.
Then \eqn{b1corr} implies that 
\be
 b_1(\kb) = b_{10} + \frac{k^2s}{\sig_1^2}b_{11}\,.
 \label{b1Fourier}
\ee
This shows that the excursion set analysis makes a prediction for how 
the \emph{Fourier}-space coefficients $b_{10}$ and $b_{11}$ should 
depend on $\nu=\delta_c/\sigma$ and $\Gamma$.  

It is remarkable that peaks theory predicts this same structure 
(constant plus $k^2$) for the linear Fourier space bias factor 
(Desjacques \etal\ 2010). Although the coefficients $b_{10}$ and 
$b_{11}$ for peaks differ from that for the excursion set halos 
studied here, the relation \eqref{b10-b11-corr} between these 
coefficients is the same.  We have checked explicitly that peaks also 
satisfy the relationships between the \emph{second} order bias 
coefficients as shown in \eqn{b20-est} (although the actual values 
of $b_{20}$, $b_{21}$ and $b_{22}$ are different), and so we expect this 
correspondence between the $k$-dependence of peak and halo-bias will 
hold for all $n$. Because this correspondence is seen in two very 
different analyses (excursion sets and peaks), there is likely to be 
a deeper reason for its existence. 

We explore this further in Appendix~\ref{app-matsubara} where we 
discuss the relation between our analysis and the work of 
Matsubara (2011) who has argued that $k$-dependent bias factors are 
generically associated with nonlocal biasing schemes.  He provides 
a number of generic results for such nonlocal bias, noting that the 
Fourier-space structure at order $n$ which can be written in terms 
of what he calls renormalized bias coefficients $c_n(\kb_1,\ldots,\kb_n)$.  
For peaks theory, 
\begin{align}
 c_n(\kb_1,\ldots,\kb_n)&= b_{n0} + b_{n1}\sum_i k_i^2 
                                 + b_{n2}\sum_{i<j} k_i^2k_j^2 + \ldots .
\label{fourierdelh-n}
\end{align}
In this case, for Gaussian initial conditions, the Hermite-weighted 
averages (with a Gaussian filter as per equation~\ref{bn-equal-cn}) 
show a structure that is identical to our excursion set predictions 
of \eqref{bnk}. More generally, our Hermite-weighting scheme provides 
a practical way of measuring integrals of Matsubara's renormalized 
bias coefficients $c_n$.

We therefore conclude that our real-space Hermite-weighted prescription 
for measuring halo bias can allow us to separate the scale-dependent 
contribution to bias as well as isolate the scale-\emph{independent} 
(peak-background split) part arising from each order $n$, which 
traditional Fourier-space measurements cannot do.
The specific results of our excursion set analysis (e.g., the 
relations between the $b_{nk}$) are then predictions that can be 
tested in more realistic settings such as $N$-body simulations.  But 
this is beyond the scope of the present work.

\section{Conclusions and discussion}
\label{conclusions}

We provided an analytic approximation for the first crossing distribution for walks with correlated steps which are constrained to pass through a specified position (equation~\ref{condfc-explicit}), and showed that it was accurate (\fig{vfvCond}).  Although this is interesting in its own right, we did not explore this further.  Rather, we used it to provide a simple analytic expression for the large scale halo bias factors (equation~\ref{bn-corr-explicit}), showing that, as a result of correlations between scales, real space measures of halo bias are scale dependent (equation~\ref{bnk} and \fig{biascompare}), but this scale dependence is best thought of as arising from $k$-dependent bias in Fourier space (Section~\ref{biasfourier}).  
Although we presented comparisons with numerical results for a specific choice of filter (Gaussian) and power spectrum ($P(k)\propto k^{-1.2}$), the results of Musso \& Sheth (2012) lead us to expect that our analytical results will be equally accurate for other filters and power spectra, including TopHat filtered $\Lambda$CDM.

For correlations which arise because of a Gaussian smoothing filter, the linear bias factor $b_1$ is a constant plus a term which is proportional to $k^2$.  This is a consequence of the fact that our analysis is based on the approximation of Musso \& Sheth (2012), which associates halos with places where the height of the smoothed field and its first derivative with respect to smoothing scale satisfy certain constraints.  If constraining the second derivative as well leads to an even more accurate model of the first crossing distribution, then this would give rise to $k^4$-dependence in the bias.  It is in this sense that $k$-dependent halo bias is part and parcel of the excursion set approach.  Such $k$-dependence will lead to stochasticity in real space measures of bias (Desjacques \& Sheth 2010); we have not pursued this further.  

We also provided an algorithm for estimating the scale-independent coefficients of the $k$-dependent bias factors from real space measurements (Section~\ref{recover}).  Although the method uses cross-correlations between the halo field and suitably transformed versions of the smoothed mass field at the same spatial position (equation~\ref{bn-crosscorr}), the bias factors it returns are independent of the scale on which this transformation is done (\fig{biasnoep}).  In particular, the coefficient of the $k$-independent part of the bias which our algorithm returns equals that associated with the peak-background split argument, even though our algorithm can be applied on scales for which the usual formulation of the peak-background split argument does not apply.  

For Gaussian fields, the transformation we advocate uses the Hermite polynomials.  Therefore, our work has an interesting connection to Szalay (1988) who noted that, instead of defining bias coefficients by writing \delh\ as a Taylor series in \delo\ as is usually done, one could have chosen to expand the mass field in Hermite polynomials.  Our analysis shows that this is indeed a fruitful way to proceed, even when the bias factors are $k$-dependent.  

There are two reasons why this is remarkable.  
First, our analysis shows that, for the excursion set model, the coefficients of the expansion in \delo\ are the {\em same} as those for the expansion in Hermite polynomials (equations~\ref{bn-crosscorr} and~\ref{bias-corr-full}).  There is no reason why this should be true in general.  And second, Szalay explicitly assumed that halo bias was `local':  \delh\ was a function of \delo\ only.  For local bias, the bias factors are $k$-independent; $k$-dependent bias factors are a signature that the bias is nonlocal (Matsubara 2011, with the $k$-dependence of peak bias discussed in Desjacques et al. 2010 being a specific example), so it is not a priori obvious that an expansion in Hermite polynomials would have been useful.  

In Appendix~B we showed why, even for nonlocally biased tracers of a Gaussian field, the Hermites are so special.  For completeness, we also provided an analysis of the general case, in which the underlying field is not necessarily Gaussian (equation~\ref{bnGeneral}).  This more general analysis may prove useful should it turn out that the primordial fluctuation field was non-Gaussian, or if one wishes to describe halo bias with respect to the nonlinear Eulerian field rather than with respect to the initial one.  

In the former case, primordial non-Gaussianity is expected to be sufficiently weak that the Edgeworth expansion can be used to provide insight into the expected modifications to halo abundances.  Since Hermites play an important role in the Edgeworth expansion, it is likely that our Hermite-based algorithm for halo bias will be useful for constraining $f_{\rm NL}$.  

Recent work has emphasized the advantages of using cross- rather than auto-correlations to estimate halo bias (Smith et al. 2007; Pollack et al. 2012).  Since $H_n$ is an $n$-th order polynomial in the mass field, one may think of our algorithm as an extension of this program:  it uses two-scale halo-mass cross-correlations at the same real-space position to extract information which is usually obtained from $n$-point statistics.  However, in addition to being simpler, our algorithm is able to estimate the bias coefficients on smaller scales than those on which the more traditional analyses $n$-point (Fourier or real-space) analyses are performed.  So we expect it to find use in analyses of halo bias in simulations, and galaxy bias in real datasets.  

For example, one can compare our prescription with traditional methods of estimating bias in real space, e.g. Manera \& Gazta\~naga (2012). Here, instead of computing averages of the matter field centered at locations of halos (as is natural in the excursion set approach), one explicitly defines a halo \emph{field} $\delh(\xb)$ smoothed on a grid of cell-size $R_0$ and uses the matter field $\delo(\xb)$ smoothed on the same grid. One then fits a polynomial of the type $\delh=b_0+b_1\delo+b_2\delo^2/2$ to a scatter plot of \delh\ vs. \delo\ using a least squares prescription. This is conceptually the same as approximating the function $\avg{\delh|\delo}$ (which is most easily seen by considering linear biasing of a Gaussian field, for which the statement is exact). This can be compared with the excursion set prediction $\avg{1+\delh|\delo} = f(s|\delo,\So)/f(s)$, and we see that the coefficients obtained from the fit will generically depend on \So. As Manera \& Gazta\~naga show, one needs to define a grid on very large scales ($R_0\gtrsim 40h^{-1}$Mpc) in order to recover scale independent bias coefficients. On the other hand, our prescription can in principle operate at much smaller scales (c.f. section~\ref{recover}) and remove this scale dependence by basically computing weighted integrals of the mean relation in the \delh-\delo\ scatter plot. A more detailed comparison with traditional techniques is complicated by the fact that we have made predictions for Lagrangian bias whereas analyses such as Manera \& Gazta\~naga's typically work in the final, Eulerian field. We leave such a comparison to future work.

In this context, it is worth noting that our algorithm is more than just a simple way of estimating the nonlinear bias coefficients $b_n$.  For example, there has been recent interest in reducing the stochasticity between the underlying mass field and that defined by the biased tracers (Hamaus et al. 2010; Cai et al. 2011).  Some of this stochasticity is due to the nonlinear nature of the bias (Hamaus et al. 2011).  Our demonstration that the nonlinear bias factors measure the amplitude of the cross-correlation function between the halo field and the Hermite-transformed mass field will simplify such analyses.

\section*{acknowledgements}
We are grateful to E. Sefusatti and M. Simonovic for discussions.  
MM is supported by ESA under the Belgian Federal PRODEX program
$\mathrm{N}^\circ 4000103071$.
This work is supported in part by NSF 0908241 and NASA NNX11A125G.

\appendix
\section{Details of calculations}
\label{app-details}
In this Appendix we sketch the proofs of various identities stated in the main text.
\subsection{Proof of \eqn{bn-crosscorr-shk}}
\label{app-shk}
To prove \eqn{bn-crosscorr-shk} for sharp-$k$ walks, it is useful to
consider the following Fourier transform relations involving the
Hermite polynomials, which follow from the definition of the $H_n$: 
\begin{align}
\frac{{\rm e}^{-x^2/2}}{\sqrt{2\pi}}H_n(x) &= \int_{-\infty}^\infty
\frac{\der k}{(2\pi)}{\rm e}^{ikx}(-ik)^n{\rm e}^{-k^2/2} \notag\\
(-ik)^n{\rm e}^{-k^2/2} &= \int_{-\infty}^\infty\frac{\der
  x}{\sqrt{2\pi}} {\rm e}^{-ikx}{\rm e}^{-x^2/2}H_n(x)\,.
\label{Hn-Fourier}
\end{align}
For the conditional first crossing distribution of \eqn{fc-LC},
we use the relation
\begin{align}
sf_{\rm u}(s|\delo,\So) &= s\left( -\frac{\p}{\p\delc}\right)
p_{\rm G}(\delo-\delc;s-\So)\,.
\label{fc-LC-rewrite}
\end{align}
Using $y_0\equiv\delo/\sqrt{\So}$ and $\nu=\delc/\sqrt{s}$ one can write
\begin{align}
&\avg{sf_{\rm u}(s|\delo,\So)H_n(\delo/\sqrt{\So})} \notag\\
&\ph{sf(s)}
= -\frac{\p}{\p\nu} \int_{-\infty}^\infty 
\frac{\der y_0}{2\pi} 
\frac{{\rm e}^{-y_0^2/2}H_n(y_0)}{\sqrt{1-\So/s}} {\rm
  e}^{-\frac{(y_0-\nu\sqrt{s/\So})^2}{2(s/\So-1)}} \notag\\
&\ph{sf(s)}
=-\left(\frac{\So}{s}\right)^{n/2}\frac{\p}{\p\nu}
\int_{-\infty}^\infty \frac{\der k}{2\pi} (-ik)^n {\rm
  e}^{ik\nu-k^2/2} \notag\\
&\ph{sf(s)}
=\left(\frac{\So}{s}\right)^{n/2} \frac1{\sqrt{2\pi}} {\rm
  e}^{-\nu^2/2} H_{n+1}(\nu)\,,
\label{fcavg-shk}
\end{align}
where the second equality follows from writing the Fourier integrals corresponding to the Hermite polynomial and the Gaussian in $(y_0-\nu\sqrt{s/\So})$, doing the integral over $y_0$ to give a Dirac delta and using this to perform one Fourier-space integral. The third equality then follows from \eqn{Hn-Fourier}. 
Together with $sf_{\rm u}(s) = (2\pi)^{-1/2}\nu\,{\rm e}^{-\nu^2/2}$, this gives the result.

\subsection{Form of bias coefficients in \eqn{bias-corr}}
\label{app-biascorr-def}
The weighted average of the distribution \eqref{fc-MS-cond} is
\begin{align}
&\avg{f(s|\delo,\So)H_n(\delo/\sqrt{\So})} \notag\\
&=\int_0^\infty\!\!\der\del^\prime\del^\prime \!\int\der\delo \,p_{\rm G}(\delo;\So)H_n\!\!\left(\frac{\delo}{\sqrt{\So}}\right) p(\delc,\del^\prime|\delo)\,.
\label{fcavg-corr-interm}
\end{align}
The product $p_{\rm G}(\delo;\So)H_n(\delo/\sqrt{\So})$ and the bivariate Gaussian $p(\delc,\del^\prime|\delo)$ (equation~\ref{condgauss}) can be expressed in terms of their Fourier transforms: i.e., we use \eqn{Hn-Fourier} and
\begin{align}
p(\delc,\del^\prime|\delo) &= \int\frac{\der^2k}{(2\pi)^2} {\rm e}^{i\kb^{\rm T}(\Del-\bar\xb)} {\rm e}^{-\frac12\kb^{\rm T}(\C-\tilde\cb) \kb}\,,
\label{condbivarGauss-Fourier}
\end{align}
with $\Del=(\delc,\del')$ and $\bar\xb$, \C\ and $\tilde\cb$ given by \eqn{condmean} and~\eqref{C2}, respectively.
The integral over \delo\ then gives a one-dimensional Dirac delta $\dir\left(k_0 - k\Sc/\So - k^\prime\epc\Sc/2s\So\right)$ where $k_0$, $k$ and $k^\prime$ are the Fourier variables corresponding to \delo, \delc\ and $\del^\prime$, respectively. Performing the $k_0$ integral gives an expression in which the contribution of the ``correction'' matrix $\tilde\cb$ exactly cancels. The result can be expressed as
\begin{align}
&\frac{1}{\So^{n/2}}
 \avg{f(s|\delo,\So)H_n(\delo/\sqrt{\So})} \notag\\
&= \left(-\frac{\Sc}{\So}\right)^{\!n} 
 \!\int_0^\infty\!\der\del^\prime\del^\prime \left(\frac{\p}{\p\delc} + \frac{\epc}{2s}\frac{\p}{\p\del^\prime}\right)^{\!n} p(\delc,\del^\prime)\,,
\label{fcavg-corr}
\end{align}
and using $\avg{1+\delh|\delo,\So}\equiv f(s|\delo,\So)/f(s)$ 
gives the result \eqref{bias-corr}.

\subsection{Taylor expansion of the conditional first crossing distribution in \eqn{fc-MS-cond-Taylor}}
\label{app-condfc-taylor}
Using \eqn{condgauss} and the shorthand notation $p_{\rm G}$ for $p_{\rm G}(\Del;\C-\tilde\cb)$ where $\Del=(\delc,\del^\prime)$ and the matrices \C\ and $\tilde\cb$ were defined in \eqn{C2}, straightforward algebra shows that 
\begin{align}
&p(\delc,\del^\prime|\delo)\notag\\ 
&= \sum_{m,n=0}^\infty
\frac{(-\delo\Sc/\So)^{m+n}}{m!n!}
\left(\frac{\p}{\p\delc}\right)^m\!
\left(\frac{\epc}{2s}\frac{\p}{\p\del^\prime}\right)^{\!n} 
p_{\rm G}\notag\\
&= \sum_{k=0}^\infty\frac{\delo^k}{k!}
\left(-\frac{\Sc}{\So}\right)^{\!k} 
\sum_{n=0}^k \binom{k}{n}
\left(\frac{\p}{\p\delc}\right)^{\!n} \!\left(
\frac{\epc}{2s}\frac{\p}{\p\del^\prime} \right)^{\!k-n}
p_{\rm G}\notag\\
&= \sum_{k=0}^\infty\frac{\delo^k}{k!}\left(-\frac{\Sc}{\So}\right)^{\!k} \left(\frac{\p}{\p\delc} +
\frac{\epc}{2s}\frac{\p}{\p\del^\prime} \right)^k 
p_{\rm G}(\Del;\C-\tilde\cb)\,.
\label{bivarGauss-Taylor}
\end{align}
Using this in the definition \eqref{fc-MS-cond} proves \eqn{fc-MS-cond-Taylor}.

\subsection{Explicit expressions for the bias coefficients in \eqn{bn-corr-explicit}}
\label{app-bias-explicit}
The explicit form of the conditional distribution \eqref{fc-MS-cond} in the limit $\tilde\cb\to0$ allows for a more convenient calculation of the bias coefficients than computing the derivatives in \eqn{bias-corr}. Using the relations \eqref{ctilto0} in \eqn{condfc-explicit} brings the conditional first crossing distribution to the form
\begin{align}
sf(s|\delo,\tilde\cb=0) 
&= \frac{{\rm e}^{-\nu^2(1-\bar\del_0\Sc\!/\So)^2\!/2}}{2\Gamma\sqrt{2\pi}}
\notag\\
&\ph{2\pi}
\times\int_0^{\infty}\frac{\der y\,y}{\sqrt{2\pi}} {\rm e}^{-\left(y-\Gamma\nu + \bar\del_0\nu_1\right)^2/2}\,,
\label{condfc-ctilto0}
\end{align}
where $\bar\del_0=\delo/\delc$ and $\nu_1=\Gamma\nu(\Sc/\So)(1-\epc)$. The Taylor expansion of this expression in powers of $\bar\del_0$ can now be used to read off the bias coefficients $b_n$ using \eqn{bias-corr-full}. The Gaussian multiplying the integral can be expanded using the definition of the Hermite polynomials $H_n(\nu)$. The following relations are useful in simplifying the integral:
\begin{align}
\int_0^\infty \der z\,z\,p_{\rm G}(z-\Gamma\nu;1) &= \frac{{\rm e}^{-\frac12\Gamma^2\nu^2}}{\sqrt{2\pi}}\left(1+A\right) \,,\notag\\
\int_0^\infty \der z\,p_{\rm G}(z-\Gamma\nu;1) &= \frac{{\rm e}^{-\frac12\Gamma^2\nu^2}}{\sqrt{2\pi}}\,\frac{A}{\Gamma\nu} \,,\notag\\
\left. \frac{\p^{n}}{\p z^{n}}[z\,p_{\rm G}(z-\Gamma\nu;1)] \right|_{z=0} &= \frac{{\rm e}^{-\frac12\Gamma^2\nu^2}}{\sqrt{2\pi}}nH_{n-1}(\Gamma\nu)\,,\notag\\
\left. \frac{\p^{n}}{\p z^{n}}p_{\rm G}(z-\Gamma\nu;1) \right|_{z=0} &= \frac{{\rm e}^{-\frac12\Gamma^2\nu^2} }{\sqrt{2\pi}}H_{n}(\Gamma\nu)\,,
\label{useful}
\end{align}
where $A$ was defined in \eqn{A-def}. Some manipulation then leads to the result quoted in \eqn{bn-corr-explicit}.

\section{Relation between Matsubara's renormalised coefficients and
  weighted averages of the matter density}
\label{app-matsubara}
Matsubara (2011) has argued that $k$-dependent bias factors are
generically associated with nonlocal biasing schemes and has provided
a number of generic results for such nonlocal bias.
In this appendix we show the connection between the
``renormalised'' coefficients $c_n(\kb_1,\ldots,\kb_n)$ defined by
him in terms of functional derivatives of the Fourier-space halo field
$\delh(\kb)$ with respect to the matter field $\del_{\kb}$, 
\be
c_n(\kb_1,\ldots,\kb_n)=(2\pi)^{3n}\int\frac{\der^3k}{(2\pi)^3}
\avg{\frac{\del^n\delh(\kb)}{\del\del_{\kb_1}\ldots \del\del_{\kb_n}}}\,,
\label{matsubara-cn}
\ee
and the real-space weighted averages of the matter density field which
we discuss in the main text. In particular, for Gaussian initial conditions, we show that the Hermite-weighted bias coefficients $b_n$ of \eqn{bn-crosscorr} are just the integrals of the $c_n$, provided one formally uses the quantity $\rhoh(\kb)$ rather than $\delh(\kb)$ in defining the $c_n$, where $\rhoh(\xb)\equiv1+\delh(\xb)$. In this case,
\begin{align}
b_n &=\frac1{\So^{n/2}}\avg{(1+\delh) H_n(\delo/\sqrt{\So})} \notag\\
   &=\frac1{\So^{n}} \int \frac{\der^3k_1}{(2\pi)^3} \ldots \frac{\der^3k_n}{(2\pi)^3}P_1\ldots P_n W_1\ldots W_n\notag\\ 
   &\ph{\int \frac{\der^3k_1}{(2\pi)^3} \ldots \frac{\der^3k_n}{(2\pi)^3}}\times c_n(\kb_1,\ldots,\kb_n)\,,
\label{bn-equal-cn}
\end{align}
where $P_i=P(k_i)$, $W_i=W(k_iR_0)$ and $\So = (2\pi)^{-3}\int\der^3kP(k)W(kR_0)^2$.

We demonstrate this in section~\ref{app-qequalp} by working in Fourier space and explicitly evaluating the integral in the second line of \eqn{bn-equal-cn}. In section~\ref{app-connected} we work in real space, repeating the calculation in field theoretic language and showing that the bias coefficients can be interpreted as connected expectation values.
This real-space calculation also shows how one might generalise our results to the case when the distribution of the matter field is not Gaussian.

\subsection{Fourier space calculation}
\label{app-qequalp}
To prove \eqn{bn-equal-cn}, note that in the definition \eqref{matsubara-cn}, the functional derivatives can be transferred to the probability density functional (which we denote as $\Cal{P}[\del_{\kb}]$),
\begin{align}
&\avg{\frac{\del^n\rhoh(\kb)}{\del\del_{\kb_1}\ldots \del\del_{\kb_n}}} \notag\\
&=\int\Cal{D}[\del_{\kb}] \Cal{P}[\del_{\kb}] \frac{\del^n\rhoh(\kb)}{\del\del_{\kb_1}\ldots \del\del_{\kb_n}} \notag\\
&=(-1)^n\int\Cal{D}[\del_{\kb}] \frac{\del^n\Cal{P}[\del_{\kb}]}{\del\del_{\kb_1}\ldots \del\del_{\kb_n}} \rhoh(\kb)\,,
\label{functionalexpecvalue}
\end{align}
where $\int\Cal{D}[\del_{\kb}]$ denotes a functional integral. Also, statistical homogeneity allows us to introduce $1={\rm e}^{i(\kb+\kb_{1\ldots n})\cdot \xb}$ where $\kb_{1\ldots n}=\kb_1+\ldots+\kb_n$ and hence write the second line of \eqref{bn-equal-cn} as
\begin{align}
&\int\Cal{D}[\del_{\kb}]\int\frac{\der^3k}{(2\pi)^3}{\rm e}^{i\kb\cdot\xb} \rhoh(\kb) \int \frac{\der^3k_1}{(2\pi)^3} \ldots \frac{\der^3k_n}{(2\pi)^3} {\rm e}^{i\kb_{1\ldots n}\cdot\xb} \notag\\
&\times \frac{W_1\ldots W_n}{\So^n}\,(-1)^n(2\pi)^{3n}P_1\ldots P_n\frac{\del^n\Cal{P}[\del_{\kb}]}{\del\del_{\kb_1}\ldots \del\del_{\kb_n}}\,.
\label{expecvalue-interm}
\end{align}
For Gaussian initial conditions, $\Cal{P}[\del_{\kb}] \propto
\exp\left[-(1/2)\int\der^3k\,\del_{\kb}\del^\ast_{\kb}/((2\pi)^3P(k))\right]$. 
The functional derivative of $\Cal{P}[\del_{\kb}]$ can then be understood
as follows (see also Matsubara 1995).
Consider the action of a single functional derivative $\del/\del\del_{\kb_i}$. When this acts on the distribution $\Cal{P}[\del_{\kb}]$, it brings down a factor $(-1)\del^\ast_{\kb_i}(2\pi)^{-3}P(k_i)^{-1}$. On the other hand, when it acts on an existing factor of $\del^\ast_{\kb_j}$, it gives a Dirac delta $\dir(\kb_i+\kb_j)$ (since $\del^\ast_{\kb_j}=\del_{-\kb_j}$). The result of $n$ derivatives on $\Cal{P}[\del_{\kb}]$ can be organised as an alternating sum over terms containing an increasing number of Dirac deltas or connections between pairs of vectors $\kb_i$, $\kb_j$. The alternation arises because each connected pair carries a minus sign. For the $n$-th derivative, the term containing $p$ connected pairs (when multiplied by $(-1)^n(2\pi)^{3n}P_1\ldots P_n$) looks like
\begin{align}
(-1)^p\bigg[& \del_{\kb_1}\ldots\del_{\kb_{n-2p}} (2\pi)^{3p} \notag\\
&\times P_{n-2p+2}\,\dir(\kb_{n-2p+1}+\kb_{n-2p+2}) \ldots 
\notag\\ &\times P_{n}\,\dir(\kb_{n-1}+\kb_{n})
+ {\rm perms.}\bigg]\,,
\label{p-pairs}
\end{align}
where ``perms.'' indicates all permutations of the vectors $\kb_j$. Since we integrate over all the $\kb_j$ with a totally symmetric prefactor ${\rm e}^{i\kb_{1\ldots n}\cdot \xb}(W_1\ldots W_n)$, all these permutations lead to identical contributions. 

The product of $(2\pi)^{3n}P_1\ldots P_n$ with the $n^{\rm th}$ derivative of $\Cal{P}[\del_{\kb}]$ therefore equals $\Cal{P}[\del_{\kb}]$ multiplied by
\begin{align}
\sum_{p=0}^{[n/2]}&(-1)^p\binom{n}{n-2p}(2p-1)!!
\,(2\pi)^{3p}\,\del_{\kb_1}\ldots\del_{\kb_{n-2p}}  \notag\\
&\times 
\prod_{j=0}^{p-1} P_{n-2j}\,\dir(\kb_{n-2j-1}+\kb_{n-2j}) \,,
\label{nth-deriv}
\end{align}
where $[n/2]$ is the floor of $n/2$ and the combinatorial factor counts the number of partitions of $n$ distinct objects into $(n-2p)$ singletons and $p$ pairs, which is precisely the coefficient of $x^{n-2p}$ in the Hermite polynomial $H_n(x)$.

On performing the integrals over $\kb_i$ in the term with $p$ connected pairs, the factors of $\del_{\kb_i}$ will contribute $(n-2p)$ powers of $\delo(\xb)$ and the Dirac deltas will contribute $p$ powers of \So. Further identifying the inverse Fourier transform of $\rhoh(\kb)$ in the first line of \eqref{expecvalue-interm}, we can write the expression in \eqref{expecvalue-interm} as
\begin{align}
&\int\Cal{D}[\del_{\kb}]\Cal{P}[\del_{\kb}]\,\rhoh(\xb) \notag\\
&\ph{\int}  \times \frac1{\So^{n/2}}\sum_{p=0}^{[n/2]}(-1)^p\binom{n}{n-2p}(2p-1)!! \left(\frac{\delo}{\sqrt{\So}}\right)^{n-2p}\notag\\
&=\frac1{\So^{n/2}}\avg{(1+\delh) H_n(\delo/\sqrt{\So})}\,,
\label{nth-deriv-FT}
\end{align}
which completes the proof.

\subsection{Real space calculation: bias as connected expectation values}
\label{app-connected}
In real space, the statement that $\rhoh(\kb)$ can be expressed in terms of the modes $\del_{\kb}$ of the matter field translates to the generic expansion
\begin{align}
  \rhoh(\xb) &= \sum_{k=0}^{\infty} \frac{1}{k!}
  \int \der^3y_1\dots\der^3y_k \, b_k(\xb-\yb_1,\dots,\xb-\yb_k) \notag\\
&\ph{\int \der^3y_1\dots\der^3y_k }  
\times \del(\yb_1)\dots\del(\yb_k)\,,
\label{app-dh-real}
\end{align}
where the $b_k$ are the coefficients of the Taylor
expansion of $\rhoh$ in powers of $\del$,
\begin{equation}
  b_k(\xb-\yb_1,\dots,\xb-\yb_k) \equiv
  \frac{\del^k\rhoh(\xb)}{\del\del(\yb_1)\dots\del\del(\yb_k)}
  \bigg\vert_{\del(\yb_i)=0}\,,
\end{equation}
which are totally symmetric in their arguments.

Each term of Equation \eqref{app-dh-real} can be considered as a vertex
with $k$ legs. The correlation function $\avg{\rhoh(\xb)\delo(\zb)}$ 
can be computed using Wick's theorem to isolate the two-point correlation 
functions connecting $\delo(\zb)$ to any of the $\del(\yb_j)$'s in the sum,
and get
\begin{align}
& \sum_{k=1}^{\infty} \frac{1}{(k-1)!}
  \int \der^3y_1\dots\der^3y_k \, b_k(\xb-\yb_1,\dots,\xb-\yb_k)\notag\\
&\ph{\sum_{k=1}^{\infty} \frac{1}{(k-1)!}} \times
  \avg{\del(\yb_1)\dots\del(\yb_{k-1})}\avg{\del(\yb_k)\delo(\zb)}\,. 
\end{align}
Since one also has
\begin{align}
  \frac{\del\rhoh(\xb)}{\del\del(\yb)}
  &= \sum_{k=1}^{\infty} \frac{1}{(k-1)!}
  \int \der^3y_1\dots\der^3y_{k-1} \notag\\ 
&\ph{ \sum_{k=1}^{\infty}}
\times b_k(\xb-\yb_1,\dots,\xb-\yb_{k-1},\xb-\yb) \notag\\
&\ph{ \sum_{k=1}^{\infty} b_k(\xb-\yb_1)}
\times   \del(\yb_1)\dots\del(\yb_{k-1}) \,,
\end{align}
then one obtains
\begin{equation}
  \avg{\rhoh(\xb)\delo(\zb)} =
  \int\der^3y \avg{\frac{\del\rhoh(\xb)}{\del\del(\yb)}}
  \avg{\del(\yb)\delo(\zb)}.
\label{dhd0c}
\end{equation}
Similarly, in order to compute any ``connected'' correlation function 
$\avg{\rhoh(\xb)\delo(\zb_1)\dots\delo(\zb_n)}_{\rm c}$ one should
retain only those terms where each of the $n$ external field is 
connected to any of the internal fields of Equation \eqref{app-dh-real}.
Since the combinatorial factors generated by the action of Wick's theorem
are the same as those obtained from differentiation, one gets
\begin{align}
&  \avg{\rhoh(\xb)\delo(\zb_1)\dots\delo(\zb_n)}_{\rm c} \notag\\
&\ph{\avg{\rhoh}}
=  \int\der^3y_1\dots\der^3y_n 
  \avg{\frac{\del^n\rhoh(\xb)}{\del\del(\yb_1)\dots\del(\yb_n)}} \notag\\
&\ph{\avg{\rhoh}\int\der^3\yb_1\dots\der^3\yb_n }
\times  \prod_{j=1}^n\avg{\del(\yb_j)\delo(\zb_j)}\,.
\label{dhd0nc}
\end{align}
Going to Fourier space one has $\avg{\del(\yb)\delo(\zb)}=
(2\pi)^{-3}\int\der^3k \,e^{i\kb\cdot(\yb-\zb)}P(k)W(kR_0)$
and $\del/\del\del_{\kb}=(2\pi)^{-3}\int\der^3y \,e^{i\kb\cdot\yb} 
(\del/\del\del(\yb))$, so that
\begin{align}
&\avg{\rhoh(\xb)\delo(\zb_1)\dots\delo(\zb_n)}_{\rm c} \notag\\
&\ph{\avg{\del_h}}
=\int\der^3k_1\dots\der^3k_n \,
  \prod_{j=1}^n \Big[e^{-i\kb_j\cdot\zb_j} P(k_j)W(k_jR_0)\Big]
  \notag\\
&\ph{\avg{\del_h}\int\der^3k_1\dots\der^3k_n}
\times  \avg{\frac{\del^n\rhoh(\xb)}{\del\del_{\kb_1}\dots\del_{\kb_n}}}\,.
\label{ncorrFourier}
\end{align}
If we write
\begin{equation}
  \avg{\frac{\del^n\rhoh(\xb)}{\del\del_{\kb_1}\dots\del_{\kb_n}}}
  \equiv \frac{e^{i(\kb_1+\dots+\kb_n)\cdot\xb}}{(2\pi)^{3n}} 
  c_n(\kb_1,\dots,\kb_n)\,,
\end{equation}
then it is not hard to see that the $c_n$ above agrees with
Matsubara's definition (equation~\ref{matsubara-cn}, with $\delh\to\rhoh$) upon requiring
statistical homogeneity, and moreover,
\begin{align}
&  \avg{\rhoh(\xb)\delo(\zb_1)\dots\delo(\zb_n)}_{\rm c} \notag\\
&\ph{\avg{}}
=\int\frac{\der^3k_1}{(2\pi)^3}\dots\frac{\der^3k_n}{(2\pi)^3} \,
  \prod_{j=1}^n \Big[ e^{i\kb_j\cdot(\xb-\zb_j)} P(k_j)W(k_jR_0)
    \Big]\notag\\
&\ph{\int\frac{\der^3k_1}{(2\pi)^3}\dots}
\times  c_n(\kb_1,\dots,\kb_n).
\end{align}
In other words, the second line of \eqn{bn-equal-cn} corresponds
to the quantity $\avg{\rhoh(\xb)\delo^n(\xb)}_{\rm c}/\So^n$, where $\So =
\avg{\delo^2(\xb)}$. 

The connected $n$-point expectation value can be recursively obtained using
\begin{align}
  \avg{\rhoh\delo} &= \avg{\rhoh\delo}_{\rm c} +  \avg{\rhoh}\avg{\delo} \notag\\
  \avg{\rhoh\delo^2} &= \avg{\rhoh\delo^2}_{\rm c}
  + 2\avg{\rhoh\delo}_{\rm c}\avg{\delo} + \avg{\rhoh}\avg{\delo^2} \notag\\
  \avg{\rhoh\delo^3} &= \avg{\rhoh\delo^3}_{\rm c}
  + 3\avg{\rhoh\delo^2}_{\rm c}\avg{\delo} \notag\\
&\ph{\avg{\rhoh\delo^3}_{\rm c}}
+ 3\avg{\rhoh\delo}_{\rm c}\avg{\delo^2}
  + \avg{\rhoh}\avg{\delo^3},
\end{align}
and in general
\begin{equation}
  \avg{\rhoh\delo^n}_{\rm c} = \avg{\rhoh\delo^n}
  -\sum_{m=0}^{n-1} \binom{n}{m}\!\avg{\rhoh\delo^{m}}_{\rm c}\!\avg{\delo^{n-m}},
 \label{bnGeneral}
\end{equation}
to remove the disconnected contributions from the average.
Since \delo\ is Gaussian-distributed, one has
$\avg{\del_0^r}=(r-1)!!\So^{r/2}$ for $r$ even and $\avg{\del_0^r}=0$
for $r$ odd; writing back $\avg{\rhoh\del_0^m}_c$ in terms of
$\avg{\rhoh\del_0^m}$ in the above expression for $m<n$, one recovers  
\begin{equation}
  \frac{\avg{\rhoh\del_0^n}_{\rm c}}{\So^{n/2}} = 
  \avg{\rhoh H_n(\del_0/\sqrt{\So})}.
\end{equation}
This therefore justifies the interpretation of the bias 
factors $b_n$ as the connected parts of the $n$-point expectation values.

Moreover, it is clear that the scale dependence of 
$\avg{\rhoh\delo^n}_{\rm c}$ comes from the presence of the $n$ mixed 
correlation functions $\avg{\del(\yb_j)\delo(\zb_j)}$ in Equation 
\eqref{dhd0nc}, introducing $n$ occurrences of the filter $W(k_jR_0)$
in Equation \eqref{ncorrFourier}. Therefore one can expect the ratio
$\avg{\rhoh\delo^n}_{\rm c}/\Sc^n$ to be approximately scale invariant.

Similar considerations hold when the distribution of the matter field 
$\delta$ is non-Gaussian. This would include both the presence of 
non-Gaussian initial conditions and non-linear gravitational evolution. In this case, each external field
$\delo(\mathbf{z_i})$ is connected to $\rho_h(\mathbf{x})$ by the full 
non-Gaussian renormalized propagator, while the coefficients 
$c_n(\kb_1,\dots,\kb_n)$ should be defined in terms of what in quantum 
field theory is usually called the 1-PI correlation function (that is, the 
sum of all the diagrams that cannot be split in two pieces by cutting one 
single line) amputated of the external legs. The bias coefficients in this case will not, in general, correspond to Hermite-weighted averages, but must be recursively constructed using \eqn{bnGeneral} (which still involves only 2-point measurements).

\label{lastpage}

\end{document}